\documentclass{article}
\usepackage{arxiv}
\usepackage[utf8]{inputenc} % allow utf-8 input
\usepackage[T1]{fontenc}    % use 8-bit T1 fonts
\usepackage{hyperref}       % hyperlinks
\usepackage{url}            % simple URL typesetting
\usepackage{booktabs}       % professional-quality tables
\usepackage{amsfonts}       % blackboard math symbols
\usepackage{nicefrac}       % compact symbols for 1/2, etc.
\usepackage{microtype}      % microtypography
\usepackage{lipsum}
\usepackage[dvipsnames]{xcolor}
\usepackage{gensymb}
\usepackage{chemfig}
\usepackage{amsmath}
\usepackage{subfigure,verbatim}
\usepackage{graphicx,subfigure,setspace,wrapfig}
\usepackage{geometry}
\usepackage{hyperref} %needed to get hyperlinks to figures
\usepackage{setspace} % needed for doublespacing
\usepackage{enumitem}
\usepackage{array,booktabs}% http://ctan.org/pkg/{array,booktabs}
\usepackage{chemfig}
\usepackage{float}
\usepackage{multirow}
\usepackage{lineno}
\usepackage[para,online,flushleft]{threeparttable}
\usepackage{amssymb}
\graphicspath{ {./images/} }
\usepackage[utf8]{inputenc}
\usepackage{amsmath}
\usepackage{graphicx}
\usepackage{lineno}
\usepackage{amssymb}
\usepackage{epstopdf}
\AppendGraphicsExtensions{.tif}
\usepackage{svg}
\usepackage{multirow}  %Adds multi-row in tables
\usepackage[export]{adjustbox}
\usepackage{soul}
\usepackage{textcomp}
\usepackage{multirow}
\usepackage{enumitem}
\usepackage{graphics}

\title{Architected Flexible Syntactic Foams: Additive Manufacturing and Reinforced Particle driven Matrix Segregation}

\author{
  Hridyesh Tewani \\
  Dept. of Civil \& Env. Engineering \\
  University of Wisconsin-Madison \\
  Madison, WI 53706 \\
  \And
  Megan Hinaus \\
  Dept. of Civil \& Env. Engineering \\
  University of Wisconsin-Madison \\
  Madison, WI 53706 \\
    \And
 Mayukh Talukdar \\
  Dept. of Civil \& Env. Engineering \\
  University of Wisconsin-Madison \\
  Madison, WI 53706 \\
    \And
  Hiroki Sone \\
  Dept. of Civil \& Env. Engineering \\
  University of Wisconsin-Madison \\
  Madison, WI 53706 \\
  \And
 Pavana Prabhakar \\
 Dept. of Mechanical Engineering \\
  Dept. of Civil \& Env. Engineering \\
  University of Wisconsin-Madison \\
  Madison, WI 53706 \\
  \texttt{pavana.prabhakar@wisc.edu} \\
}

\date{ }

\begin{document}
\maketitle
\begin{abstract}

Polymer syntactic foams are transforming materials that will shape the future of next-generation aerospace and marine structures. When manufactured using traditional processes, like compression molding, syntactic foams consist of a solid continuous polymer matrix reinforced with stiff hollow particles. However, polymer matrix segregation can be achieved during the selective laser sintering process with thermoplastic polyurethane (TPU). It is uncertain what role hollow particles play in forming this matrix segregation and its impact on the corresponding mechanical properties of syntactic foams. We show that the size of the hollow particles controls the internal microscale morphology of matrix segregation, leading to counter-intuitive macroscale mechanical responses. Particles with diameters greater than the gaps between the cell walls of the segregated matrix get lodged between and in the walls, bridging the gaps in the segregated matrix and increasing the stiffness of syntactic foams. In contrast, particles with smaller diameters with higher particle crushing strength get lodged only inside the cell walls of the segregated matrix, resulting in higher densification stresses (energy absorption). We show that stiffness and densification can be tuned while enabling lightweight syntactic foams. These novel discoveries will aid in facilitating functional and lightweight syntactic foams for cores in sandwich structures.

%Polymer syntactic foams are lightweight composites made of hollow particles reinforced in a polymer matrix, such as glass micro-balloons (GMBs) or cenospheres. This paper demonstrates for the first time that the size of the GMB controls the interior microscale architecture of syntactic foams with matrix segregation achieved through selective laser sintering. This resulted in perplexing macroscale mechanical responses. GMBs with a diameter larger than the gaps between the segregated matrix's cell walls get lodged between and in the cell walls, whereas those with a diameter smaller than the gaps are likely to get lodged inside the cell walls. As a result, larger particles have a significant effect on the stiffness of syntactic foams, while smaller particles don't have much of an effect. Larger particles with lower crushing strength lowered the densification stress of the foam. In contrast, the foam with smaller particles (higher crushing strength) behaved similarly to pure TPU but with substantially less weight. Overall, we show that combining reinforcement parameters with print parameters enables the design and fabrication of hierarchically customized structures and functional features in 3D-printed syntactic foams. These discoveries can be used to develop and design lightweight cores for sandwich structures in marine and aerospace applications.
\end{abstract}


\section{Introduction}\label{intro}

Syntactic foams are composite materials with hollow thin-walled fillers blended in a continuous polymeric \cite{Gupta2004, Bharath2020,BharathKumar2016,Patil2019,prabhakar2022densification}, metallic \cite{Peroni2014}, and ceramic \cite{Li2008} matrices. Incorporating these hollow fillers makes these foams inherently lightweight and can help achieve tailorable mechanical properties \cite{Singh2018a,Singh2019}. Due to their inherent buoyant behavior, these foams see wide applications in the marine and aerospace industries \cite{Shahapurkar2018}. Compared to other types of composite materials with reinforcements ranging from continuous fibers, carbon nanotubes, and short fibers, manufacturing particle-reinforced composites offers the advantage of low-cost production. In this paper, we focus on the additive manufacturing of lightweight flexible syntactic foam composites with a segregated matrix system. Particularly, we aim to understand the influence of the Selective Laser Sintering (SLS) process on the polymer matrix distribution in the presence of Glass Micro-Balloon (GMB) particles.

Polymeric syntactic foams have been traditionally fabricated with Glass Micro-Balloons in a thermoplastic matrix using injection molding \cite{BharathKumar2016} and compression molding \cite{Jayavardhan2017}. Kumar et al. \cite{BharathKumar2016} fabricated cenosphere/high-density polyethylene (HDPE) syntactic foams using an industrial-scale injection molding technique. They suggested optimized parameters to minimize the fracture of cenosphere particles and promote the proper mixing of particles in the HDPE matrix. They observed that increasing the cenosphere content in the HDPE matrix resulted in an increase in the tensile moduli values, but the tensile strength decreased. Surface modification of the cenosphere particles reduced the extent of the decrease in the strength of the syntactic foams. Jayavardhan et al. \cite{Jayavardhan2017} proposed a compression molding technique to manufacture Glass Microballoon (GMB)/high-density polyethylene syntactic foams. They conducted an experimental study to elucidate the impact of volume fractions of GMBs on tensile and flexural properties. They observed that the specific moduli increased for the foams with the increase in volume fraction of GMBs, but the specific strength remained comparable to neat HDPE. Although conventional manufacturing techniques, such as compression and injection molding, offer the advantages of producing syntactic foams rapidly, they can be cost-intensive and can potentially damage the fillers. At lower volume fractions, the particles also tend to coagulate \cite{Singh2019}. Additionally, employing conventional manufacturing techniques can be arduous and cost extensive to produce parts with complex shapes due to the need for complex molds. 

%Research has been conducted to evaluate the effect of GMB parameters such as particle size \cite{Yang2004} and wall thickness \cite{Gupta2004} on the mechanical response of syntactic foams. Cao et al. \cite{Cao} investigated numerically and experimentally the response of GMB reinforced polymeric syntactic foam under quasi-static and dynamic compression loading. They inferred that the yield strength increased and the shock-wave velocity decreased with an increase in the GMB content. Kuo et al. \cite{Kuo2005} also investigated ceramic particles to reinforce continuous polyether-ketone (PEEK) matrix systems to improve hardness, modulus, and tensile strength compared to unfilled PEEK. Substantial studies have been performed to manufacture hollow particle-filled polyurethane syntactic foams and understand their mechanical response under quasi-static compression loading \cite{YOUSAF2020107764,Curd2021} and dynamic impact loading \cite{Chalivendra2003}. Chalivendra et al. \cite{Chalivendra2003} experimentally investigated the fracture toughness of cenosphere filled thermoset polyurethane syntactic foams and proposed a predictive model to estimate the toughness values of the foams. 

Additive Manufacturing (AM), a manufacturing technique that works on the principle of additive design - constructing 3D CAD geometries layer by layer - can be employed to overcome the manufacturing issues mentioned above that can occur during conventional techniques \cite{Xu2021}. Research has been performed to characterize GMB reinforced high-density polyethylene (HDPE) syntactic foams manufactured using an extrusion-based AM technique, Fused-Filament Fabrication (FFF) \cite{Singh2018,Singh2018a,Patil2019,Bharath2020}. The addition of GMBs to the high-density polyethylene (HDPE) matrix was shown to increase the elastic modulus and yield strength under quasi-static compressive loading. Bonthu et al. \cite{Bonthu2020} have also identified difficulties that must be resolved in order to produce high-quality syntactic foams using the FFF production technology. Research has been conducted to overcome the limitations of the FFF process by using 3D printing techniques with thermosets in which syntactic foams are printed directly without the need to manufacture individual rasters \cite{Nawafleh2020}. These 3D printing techniques have been in the spotlight due to the low-cost production of complex-shaped syntactic foams; however, they require complicated support structures. Selective Laser Sintering (SLS) is a powder-based AM technology in which powder is deposited layer by layer. A high-energy laser beam is then used to coalesce/sinter the powder to generate a three-dimensional geometry. The unsintered powder on the print bed serves as the support structure, which can be easily removed by blowing air in the post-processing stage \cite{wegner2021introduction,greiner2017selective}, which eliminates the requirement of support structures to print complex parts. SLS is being widely used in various areas and sees applications in the biomedical \cite{Charoo2020,song2021porous}, aerospace \cite{lv2020polyetherimide}, automotive \cite{storch2003selective}, and defense \cite{chen2018high} industries. 

Currently, several thermoplastic polymer powders for SLS are commercially accessible, including polyamide 12 (PA12), polyamide 6 (PA6), polyamide 11 (PA11), thermoplastic urethane (TPU), polypropylene (PP), and polyethylene (PE) \cite{Schmid2015}. Research has been conducted to manufacture periodically patterned and topologically optimized lattice structures with pure PA12 \cite{Maskery2018,Yuan2019} and TPU \cite{Yuan2017} polymer systems to improve the energy absorption of the structures. Furthermore, SLS has been employed to manufacture TPU and PA11 composites with carbon nanotubes \cite{Yuan2016,li2017selective,Bai2014,Zhuang2020,Zhuang}, and graphene \cite{Ronca2019}. In recent times, multiple studies have been conducted on composite foams with SLS to reduce weight, enhance mechanical properties \cite{Hon2003,Mazzoli2007,zhu2016novel}, improve thermal stability \cite{Lao}, manufacture biocompatible composites for medical devices \cite{Tan2005}, and improve electrical conductivity \cite{Athreya2010,Chen2010,gaikwad2013electrical,Qi2017}, compared to pure polymer. These studies on the use of polymer blends in SLS show that micro- and nano-sized particles can be used to alter the properties of polymer parts that have been sintered. In recent years, SLS has also been in focus to produce syntactic foams by incorporating GMBs \cite{Ozbay2022, OzbayKsasoz2022} and ceramic particles \cite{Deckers2013,Tang2011,Kenzari2012} in continuous PA 12 matrix, a relatively rigid plastic compared to flexible TPU. In addition, Mousa \cite{mousa2014effects} researched avenues to improve the bonding of the GMBs by surface modifications of the microbubbles, and their influence on the tensile strength, modulus, and impact strength of the GMB-filled PA12 syntactic foams. They concluded that the tensile properties improved with increasing the GMB content (treated/untreated), but the impact strength and ductility of the foams reduced. Cano et al. \cite{cano2018effect} investigated the effect of GMBs incorporated into the PA12 matrix on the fracture toughness of the sintered parts at various temperatures. They showed that inadequate bonding between the matrix and the particles decreased the resistance to fracture. Although PA12 syntactic foams have been the subject of a significant amount of research, a relatively small number of studies have been conducted on TPU syntactic foams utilizing SLS. 
%Experimentally and theoretically, Chung and Das \cite{chung2006processing} explored the mechanical response and fabrication of PA11 syntactic foam systems filled with GMBs. They concluded that the tensile and compressive moduli increased as the GMB content increased, however, the strain to failure and yield values decreased.  

Although research has been conducted on SLS-fabricated syntactic foams with a continuous matrix system surrounding the reinforcing particles, their {\bf mechanics with a segregated matrix system} have not been thoroughly investigated. In recent studies on SLS printing with TPU \cite{li2017selective,Yuan2017,Pan2020,Toncheva2021}, it was observed that the TPU matrix forms a controlled segregated structure at the microscale, which is a discontinuous matrix system, in the sintered parts. Therefore, it is essential to comprehend the mechanics of this type of syntactic foam in which the particles are incorporated into a segregated matrix system. 

In this study, we elucidate how additive manufacturing parameters can be coupled with GMB parameters to achieve the desired mechanical response or to tune the mechanical response of syntactic foams with the segregated TPU matrix. To that end, we present an SLS-based manufacturing method for producing multi-scale architected syntactic foams with segregated TPU matrix systems containing different grades of GMBs at varied volume fractions. First, we conducted an experimental parametric study to determine the optimal print parameters for the production of our syntactic foams. Then, we determined the effects of the GMB characteristics on the mechanical response of these foams by varying the volume fractions of various classes of GMBs with varying particle size distributions. {\bf In contrast to well-studied existing syntactic foams with continuous matrix systems, in segregated matrix systems, the particles themselves can be integrated in a variety of ways, influencing the mechanical response.} Finally, we printed syntactic foams with macroscale architectures consisting of struts with microscale architecture as a result of GMBs in the segregated matrix. We examined the effect of this manipulation of microscale and macroscale architectures on the compressive response of architected syntactic foams.

%Finally, we suggest a method for incorporating print parameters into the existing theoretical micro-mechanical models for particle composites in order to account for the segregated matrix structure.

%Therefore, it is necessary to understand the mechanics of this category of syntactic foams in which the particles would be incorporated in a segregated matrix system. To this end, in this work, we present a manufacturing technique to manufacture architected TPU syntactic foams with segregated matrix systems with various grades of GMBs at different volume fractions using the SLS process. First, we performed a parametric study to finalize the print parameters to manufacture our syntactic foams. Then, we chose different grades of GMBs having different particle size distributions and varied their volume fractions to understand the effect of GMB parameters on the mechanical response of these foams. Unlike existing syntactic foams that are extensively studied, the particles themselves can be incorporated in different manners, affecting the mechanical response. In this work, we also elucidate how the printing parameters can be coupled with the GMB parameters to influence the mechanical response from the printed syntactic foams. Finally, we update the existing theoretical micro-mechanical models for particulate composites to account for the segregated matrix system and propose a way to incorporate print parameters into the model.

\section{Motivation}\label{motiv}

\begin{figure}[h!]
\centering
\includegraphics[width=0.8\textwidth]{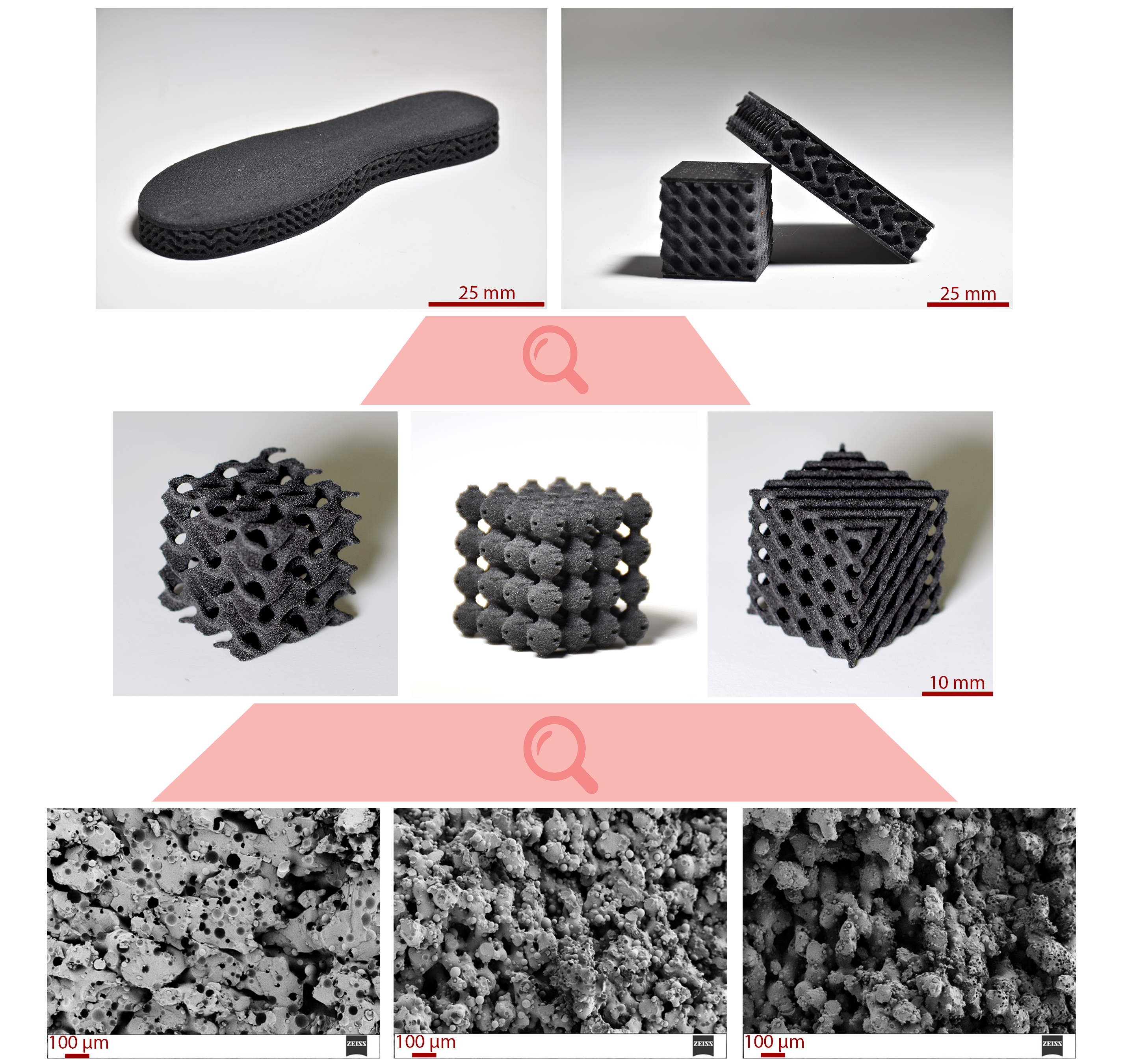}
\caption{Diverse applications of flexible syntactic foams in the sports industry (shoe sole) and aerospace industry (core for sandwich structure) with tunable macroscale and microscale architectures.}
\label{img:demon}
\end{figure}

Our overarching goal is to develop a novel manufacturing approach that combines architected hierarchy at the macroscale with design at the microscale by combining different grades of GMBs in a segregated TPU matrix. A demonstration of design ideas from the current work on flexible syntactic foams that connects microscale and macroscale architectures to diverse applications of in the sports industry (shoe sole) and aerospace industry (core for
sandwich structure) is illustrated in \textbf{Figure} \mbox{\ref{img:demon}}.

The segregated matrix can impact how the reinforcing particles get lodged in the system compared to conventional syntactic foam systems that use continuous matrix systems. As a result, understanding the deformation mechanics and corresponding mechanical performance of these syntactic foams with a segregated matrix is critical. 

In this paper, we seek to answer the following fundamental questions:

\begin{itemize}
    \item What are the effects of print parameters on the microscale morphology and mechanical performance of 3D-printed syntactic foams? %In order to comprehend these effects, we aim to achieve the required microscale hierarchy. 
    \item What is the role of GMBs in the evolution of AM-induced matrix segregation within syntactic foams?
    \item How can the GMB and print parameters be modified in unison to tune the properties, such as stiffness and densification, of additively manufactured syntactic foams?
    \item By designing and fabricating architected syntactic foams, can we tailor the macroscale mechanical properties through a multi-scale hierarchy from micro to macroscale? %This multi-scale hierarchical architecture in the syntactic foam would allow us to produce them with tailored mechanical properties.  
\end{itemize}

%%%%%%%%%
\section{Methodology}\label{method}

\begin{figure}[h!]
\centering
\includegraphics[width=0.9\textwidth]{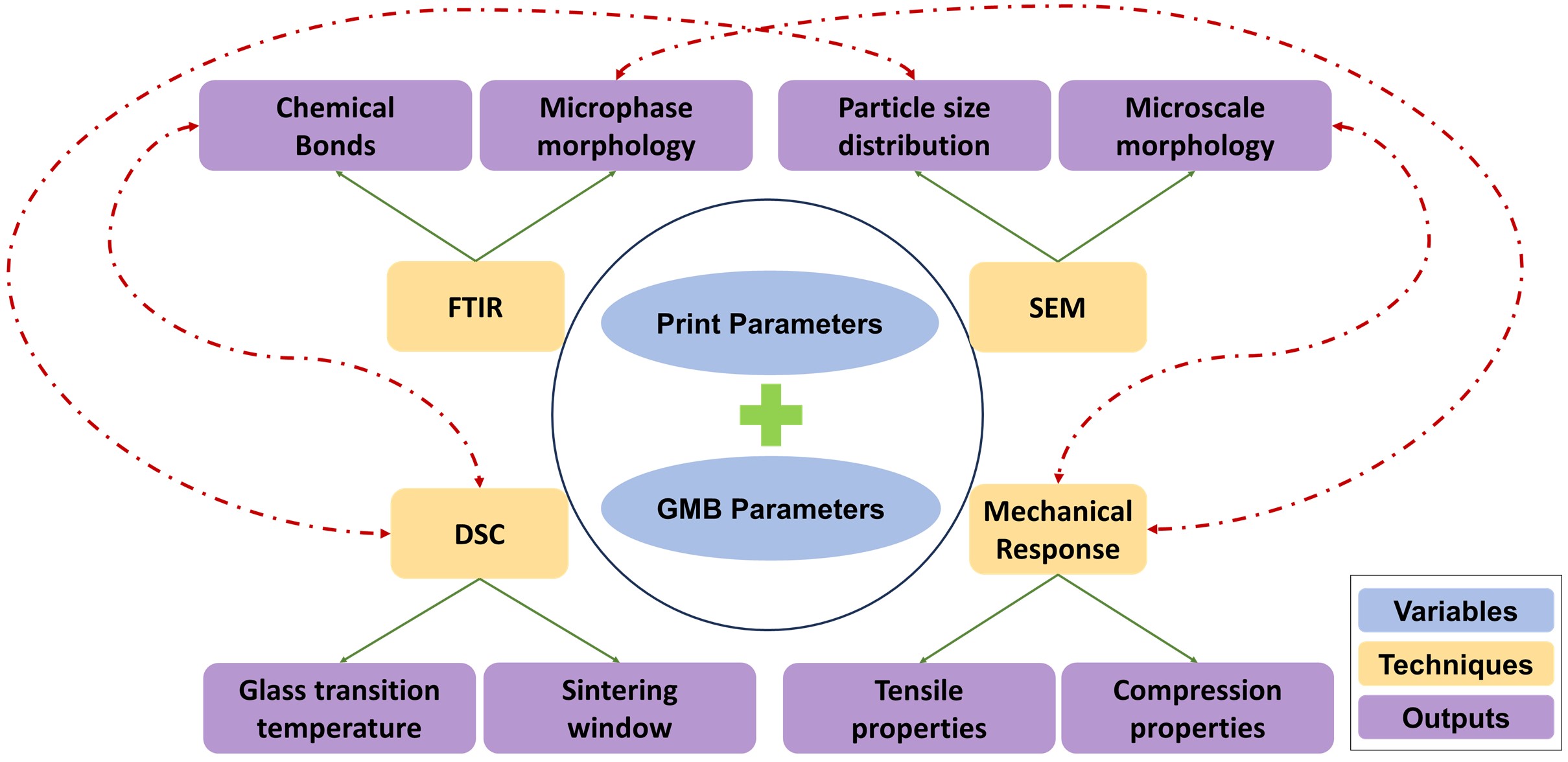}
\caption{An overview of the methodologies with interconnections between the variables, techniques, and outputs.}
\label{img:contmap}
\end{figure}

In this section, we will describe the properties of the constituent materials used for manufacturing the syntactic foams, followed by details of the SLS process and different techniques used to characterize the constituent materials. Then, we will describe the mechanical test procedures performed in compliance with the ASTM standards. An overview of the methodology section with interconnected methods with expected outcomes is summarized in \textbf{Figure} \ref{img:contmap}.  

%%%
\subsection{Materials} \label{Mat}
We procured thermoplastic polyurethane (TPU) powder from Sinterit (Product Name: Flexa Grey; particle size between 20 and 120 µm) to manufacture the syntactic foams. We chose 3M K20, 3M K46-HS, and 3M im30k grades of GMBs to create powder blends consisting of TPU powder and different volume fractions of GMBs. The properties of the constituent materials are summarized in \textbf{Table} \ref{tab:mat_prop_TPU} and \textbf{Table} \ref{tab:mat_prop_GMB}. 

\begin{table}[h!]
\centering
\renewcommand{\arraystretch}{1.2}
\caption{Properties of TPU powder (Flexa Gray from Sinterit) \cite{Sinterit2010}}
\resizebox{\columnwidth}{!}{%
\begin{tabular}{lcccccc}
\hline
\textbf{Material Property} & \textbf{Tensile Strength }   & \textbf{Elongation at Break }      & \textbf{Shore A Hardness}       & \textbf{Melting Point}   & \textbf{Softening Point }   & \textbf{Granulation }  \\
 & \textbf{(MPa)}   & \textbf{(\%)}      &        & \textbf{ (\degree C)}   & \textbf{(\degree C)}   & \textbf{($\mu$m)}  \\
\hline
Value    & 3.7 & 136 & 70 & 160 & 67.6 & 20-105   \\ \hline
\end{tabular}%
}
\label{tab:mat_prop_TPU}
\end{table}

\begin{table}[h!]
\centering
\renewcommand{\arraystretch}{1.2}
\caption{Properties of GMBs \cite{3M2013}\cite{3MAdvancedMaterialsDivision2017} (* foam representation is given for volume fraction = 20 \%)}
\resizebox{\columnwidth}{!}{%
\begin{tabular}{lcccccc}
\hline
\textbf{Grade}  & \textbf{Test Pressure )}  & \textbf{True Density } & \textbf{Particle Size - D50} & \textbf{Particle Representation} & \textbf{Foam Representation} & GMB comments\\
  & \textbf{(MPa)}  & \textbf{(g/cc)} & \textbf{($\mu$m)} &  &  & \\\hline
3M - K20  & 3.45 & 0.2 & 60  & GM60 & SF60-20 & Large particles\\ 
3M - K42HS  & 51.71 & 0.46 & 22  & GM22 & SF22-20 & Medium particles \\ 
3M - iM30k  & 193.05 & 0.60 & 15.3  & GM15 & SF15-20 & Small particles\\ \hline
\end{tabular}%
}
\label{tab:mat_prop_GMB}
\end{table}

%%%
\subsection{Manufacturing} \label{Manuf}

In this section, we will discuss the manufacturing of pure TPU and GMB-reinforced TPU syntactic foams using the SLS technique.  

\subsubsection{SLS Printing}
We manufactured the syntactic foams using a Lisa 3D printer from Sinterit \cite{Sinteritsp.zo.o.2022}, which is a desktop-based SLS printer equipped with an IR laser diode of 5W and a wavelength of 808 nm. A general description of the SLS printing process is provided in Section~\ref{app:SLSprocess}. We used pure TPU powder and TPU/GMB blends with three different volume fractions of GMBs – 20\%, 40\%, and 60\% to additively manufacture pure TPU and syntactic foams, respectively. For example, to prepare a mix with a 20\% volume fraction of GMBs, 800 ml of TPU powder and 200 ml of GMBs were measured. The mixture was then loaded into a V-shaped mixer (Power = 110 V and capacity = 1.2L), and the blend was first mixed at 30V for five minutes followed by mixing at 70V for another three minutes. To check the quality of mixing in the blend, we obtained SEM images of the powder blends; we have shown and discussed these images in detail in Section \ref{PowMixSEM}. The same mixing process was followed for all volume fractions and GMB types. 

%\begin{figure}[h!]
%    \centering
%    \includegraphics[width = 0.6\textwidth]{Images/SLS Process.jpg}
%    \caption{Illustration of SLS process.}
%    \label{img:ManProc}
%\end{figure}

%%%
\subsubsection{Print Parameters} \label{EffPP}
Identifying the optimal parameters to sinter the polymer effectively is challenging for the SLS process. Therefore, we varied the print parameters to understand their contributions to the morphology and the mechanical response of the printed foams.

%, and how these print parameters can be incorporated into the theoretical model. %During the sintering process, it is necessary for the polymer particles to undergo complete coalescence, and have good adhesion with previously sintered layers. In this work, we varied the laser power ratio and layer height to understand the effect of those parameters on the sintered polymeric parts.

\paragraph{Laser Power Ratio} \label{EffLPR}

In the Sinterit Lisa printer, the laser power ratio is controlled by two parameters: i) laser power supplied and ii) scanning speed, which affect the effective energy supplied to sinter the powder. The laser power ratio (LPR) is directly related to the laser power and inversely related to the scanning speed.
As the laser power directly increases the energy density supplied to the polymer powder, this increases the depth of the melt pool, as illustrated in \textbf{Figure} \ref{img:LPRIll}. Whereas increasing the scanning speed, decreases the dwell time, which decreases the energy supplied, and vice versa. In the Sinterit Lisa printer \mbox{\cite{Sinteritsp.zo.o.2022}}, the laser power is fixed at 5W. Therefore, by changing the LPR, we effectively vary the scanning speed while holding the power at 5W. In the current study, we used laser power ratios of 0.75, 1, 1.5, and 2, which resulted in print speeds of 83 mm/s, 56 mm/s, 37 mm/s, and 28 mm/s, respectively. When we increase the laser power ratio, it can potentially affect the micro-phase morphology in the TPU that impacts the mechanical response - this will be discussed in further detail in Section~\ref{FTIRres}.  

\begin{comment}
\begin{table}[h!]
\centering
\caption{Summary of print speeds for laser power ratios of 0.75, 1.0, 1.5, and 2.0.}
\resizebox{\columnwidth}{!}{%
\begin{tabular}{lccc}
\hline
\textbf{Laser Power Ratio (LPR)} & \textbf{Print Length(mm)}   & \textbf{Print Time (minutes)}      & \textbf{Print Speed (mm/s)}         \\ \hline
0.75    & 200000 & 4 & 83.33    \\ 
1.00    & 200000 & 6 & 55.55    \\
1.50    & 200000 & 12 & 37.04    \\
2.00    & 200000 & 18 & 27.77    \\ \hline
\end{tabular}%
}
\label{tab:printspeed}
\end{table}
\end{comment}

%To anticipate the mechanical properties of these foams, it would be necessary to include the effect of the laser power ratio in the theoretical model. %We observed that when we increased the laser power ratio, the density of the foams would increase which implies that enhancing this factor will increase the effective density of the foam - \textbf{$\rho_{fs}$}. 

\paragraph{Layer Height} \label{EffLH}

The height of the individual layer that will be placed on the print bed for each sintering phase is defined by the layer height parameter. Since SLS is a layer-based AM process, increasing the height of each layer could cause a staircase effect, even though it can shorten the print time. Furthermore, for the same provided energy density, a higher layer height lowers the bonding between individual print layers, thus compromising mechanical performance. In this work, we vary the layer height between 175 {$\mu{m}$} to 75 {$\mu{m}$} to see how it influences the mechanical response.  

%As a result, a factor to account for this print parameter can be included in the theoretical model. 

%%%
\subsection{Materials Characterization} \label{Char}
%We performed in-depth materials characterization to highlight the impact of print parameters and material compositions (for TPU/GMB blends) on the morphology, density, porosity, thermal properties, and potential degradation of the materials. These include 1) Scanning Electron Microscopy (SEM) to obtain the particle size distributions and microscale morphologies of TPU powder and TPU syntactic foams, 2) Fourier Transform Infrared (FTIR) spectroscopy to understand the chemical compositions of the constituents to manufacture the syntactic foams, and 3) Porosity Measurements to measure the porosity values of the pure and GMB included TPU syntactic foams. Details of these characterizations is provided in \ref{appendix_materials_charac}.

% Other option for Material Characterization

To evaluate the properties of the powder blends and syntactic foams, we used a variety of characterization techniques. Using scanning electron microscopy (SEM), we determined the particle size distributions and microscale morphologies of TPU powder and GMBs. Using SEM, we were also able to examine the microstructure and failure morphologies of the 3D-printed TPU and TPU/GMB foams. Fourier Transform Infrared (FTIR) spectroscopy was essential for identifying the chemical compositions of the foam production constituents. Using FTIR, we analyzed the spectroscopic properties of TPU powder and TPU/GMB composites, as well as chemical changes during the sintering process and the impact of print parameters on foam composition. Using Differential Scanning Calorimetry (DSC), we were able to evaluate the thermal properties of the TPU and determine the optimal sintering window. Using a helium porosimeter, we determined the porosity values of our pure TPU foam and TPU syntactic foams containing GMBs. In this study, we also used the porosimeter to comprehend the porosities of uncompressed and compressed TPU syntactic foams. These characterizations are further described in detail in \ref{appendix_materials_charac}.

%%%

\subsection{Mechanical Testing} \label{MechTest}
\subsubsection{Tensile testing}
We carried out uniaxial tensile tests of the printed foams in compliance with ASTM 638 \cite{ASTM638} on the MTS universal testing instrument at the Structures and Materials Testing Laboratory at the UW Madison with a load cell capacity of 250N. Type IV sample was chosen to get tested and loaded at a 50 mm/min cross-head speed. This cross-head speed was chosen such that the test completion time stays between 1 and 5 minutes. Since TPU samples typically display failures at very high elongation, an Epsilon One optical extensometer was used to obtain the engineering strains. To understand the individual effect of print parameters on tensile performance, samples were printed with different laser power ratios and layer heights. In addition, samples with various GMB volume fractions produced using the finalized print parameters were examined under tensile stress.

\subsubsection{Compression testing}

We conducted uniaxial compression tests on TPU foam samples on the ADMET 2613 tabletop frame equipped with a load cell capacity of 50 kN. The ASTM D1621 \cite{ASTM1621} standard for compression testing of plastics was used for these tests, and the sample size was chosen as 25mm x 25mm x 25mm for the cube and all architected designs. Samples were loaded under uniaxial compression at a loading rate of 2.5 mm/min (10\% of height per minute) to achieve strain values of 20\%, 30\%, and 50\%, and a preload of 1N was used. We subjected the compressed samples to a second loading cycle one week after the initial loading cycle to examine the cyclic behavior under compression. All samples were loaded to 50\% strain values for the second cycle.

\section{Results and Discussion}\label{res}
In this section, we discuss the characterization results for the constituent powders followed by an evaluation of the printed foam morphology. In addition, the impact of the print settings and GMB parameters on the mechanical performance of the foam is discussed. Finally, we elucidate the effect of GMB volume fraction and size on the mechanical response of these printed syntactic foams. %The summary of this section is shown in \textbf{Figure} \ref{img:ressumm}.

\subsection{Characterization} \label{CharRes}
We used SEM to characterize the particle size distribution of the constituent powders and understand the distribution of the GMB inclusions in the powder blends. SEM images were also used to elucidate the effects of adding different GMBs on the morphology of the printed foams.

%%%

\begin{comment}
\begin{figure}[h!]
\centering
\subfigure[]{
\includegraphics[width=0.20\textwidth]{Images/Pow1.jpg}
}
\centering
\subfigure[]{
\includegraphics[width=0.20\textwidth]{Images/Pow2.jpg}
}
\subfigure[]{
\includegraphics[width=0.20\textwidth]{Images/Pow3.jpg}
}
\subfigure[]{
\includegraphics[width=0.20\textwidth]{Images/Pow4.jpg}
}
\caption{Powder morphology of the (a) TPU powder; (b) 3M-K20 GMB; (c) 3M-K46 GMB; and (d) 3M-iM30k GMB}
\label{img:Pow-SEM}
\end{figure}
\end{comment}

\subsubsection{Powder blend morphology}\label{PowMixSEM}
\begin{figure}[h!]
    \resizebox{1\textwidth}{!}{
    \begin{tabular}{|c|c|c|}
    \hline
   \textbf{TPU/GM60-20} & \textbf{TPU/GM22-20} & \textbf{TPU/GM15-20}  \\
        \hline
 \includegraphics[width=0.30\textwidth]{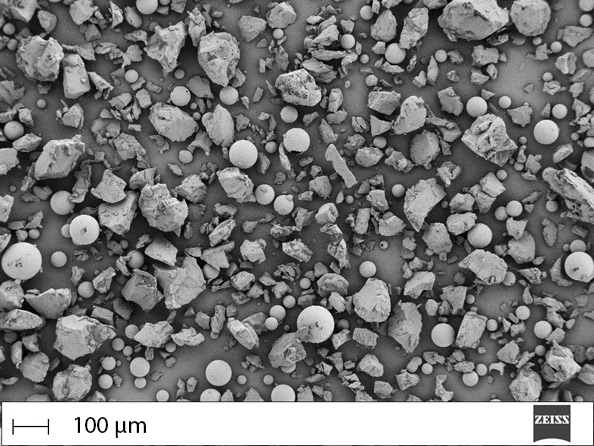} (a) &  \includegraphics[width=0.30\textwidth]{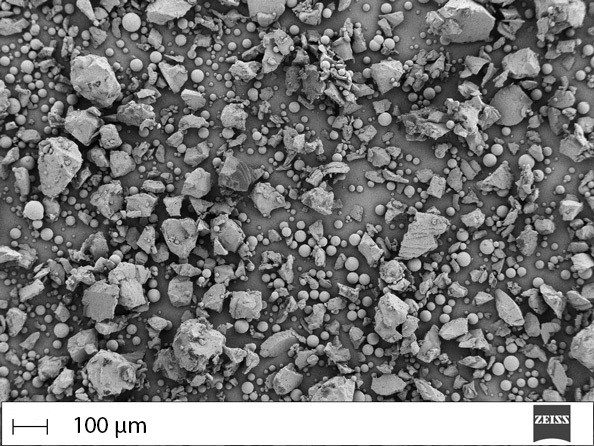} (b)
& \includegraphics[width=0.30\textwidth]{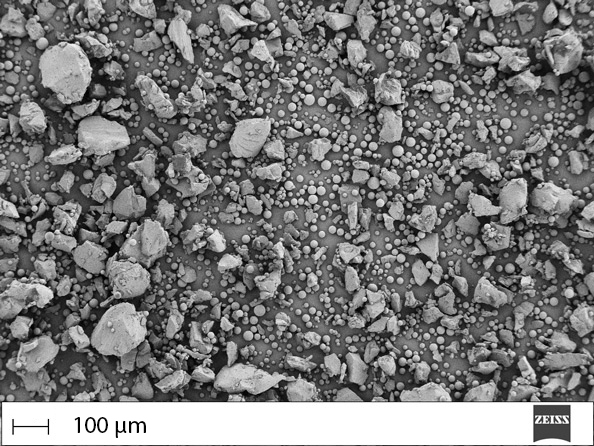} (c) \\   
        \hline

    \end{tabular} }
\caption{SEM images (E.H.T. = 3kV and Signal = SE2) of (a) TPU/GM60-20, (b) TPU/GM22-20, and (c) TPU/GM15-20 powder blends.}
\label{img:pow_mix}
\end{figure}

\textbf{Figure} \ref{img:pow_mix} shows the SEM images of TPU/GMB blends with the three sizes of GMBs mixed at a volume fraction of 20\%. We see that the polymer powder has an irregular shape, whereas all GMBs are perfectly spherical. Moreover, the smaller GMBs exhibited a higher GMB particle density per unit area than the larger GMBs. As the TPU powder and GMBs have distinct energy absorption characteristics, the GMB shape and distribution in the polymer blend have a substantial impact on the energy absorbed by the polymer powder in the mix. Smooth spherical shapes of the GMBs are expected to create less hindrance in the path of the laser, and hence, the polymer blends absorbed more energy \cite{OzbayKsasoz2022}. However, this effect can be counteracted by the hindrance caused by a higher particle density as we reduce the particle size for a fixed volume fraction. This implies that as particle density increases, we need to provide more energy density to the system. For this study, we have kept the supplied energy density constant for all TPU/GMB blends.

%Due to the smooth spherical shapes of the GMBs, they are expected to create less hindrance in the path of the laser, and hence, the polymer blends would absorb more energy \cite{OzbayKsasoz2022}. Due to their reduced tortuous shape, GMBs transmitted more energy in polymer blends, but the effective supplied energy to the polymer powder to be sintered decreased. This decrease in provided energy would be greater for mixtures containing smaller GMBs due to the larger GMB particle density per unit area.

\begin{comment}
\begin{figure}[h!]
\centering
\subfigure[]{
\includegraphics[width=0.30\textwidth]{Images/PowK20.jpg}
}
\centering
\subfigure[]{
\includegraphics[width=0.30\textwidth]{Images/PowK46.jpg}
}
\centering
\subfigure[]{
\includegraphics[width=0.30\textwidth]{Images/PowiM30k.jpg}
}
\caption{TPU and GMB distribution for the (a) TPU/GM60-20, (b) TPU/GM22-20, and (c) TPU/GM15-20 powder blends}
\label{img:pow_mix}
\end{figure}
\end{comment}

\subsubsection{FTIR Spectroscopic Analysis}\label{FTIRres}

\begin{figure}[h!]
\centering
\subfigure[]{
\includegraphics[width=0.45\textwidth]{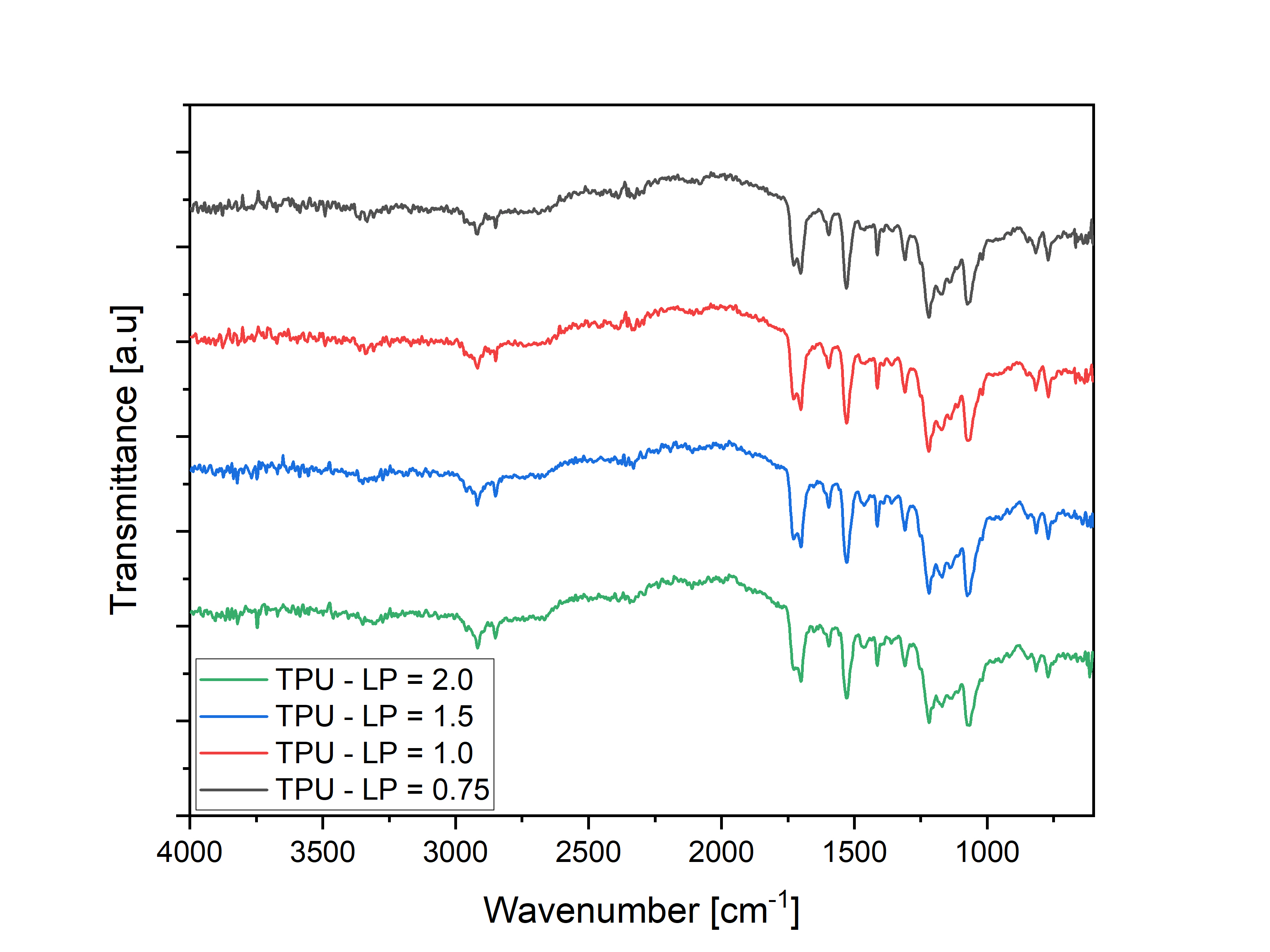}
}
\centering
\subfigure[]{
\includegraphics[width=0.45\textwidth]{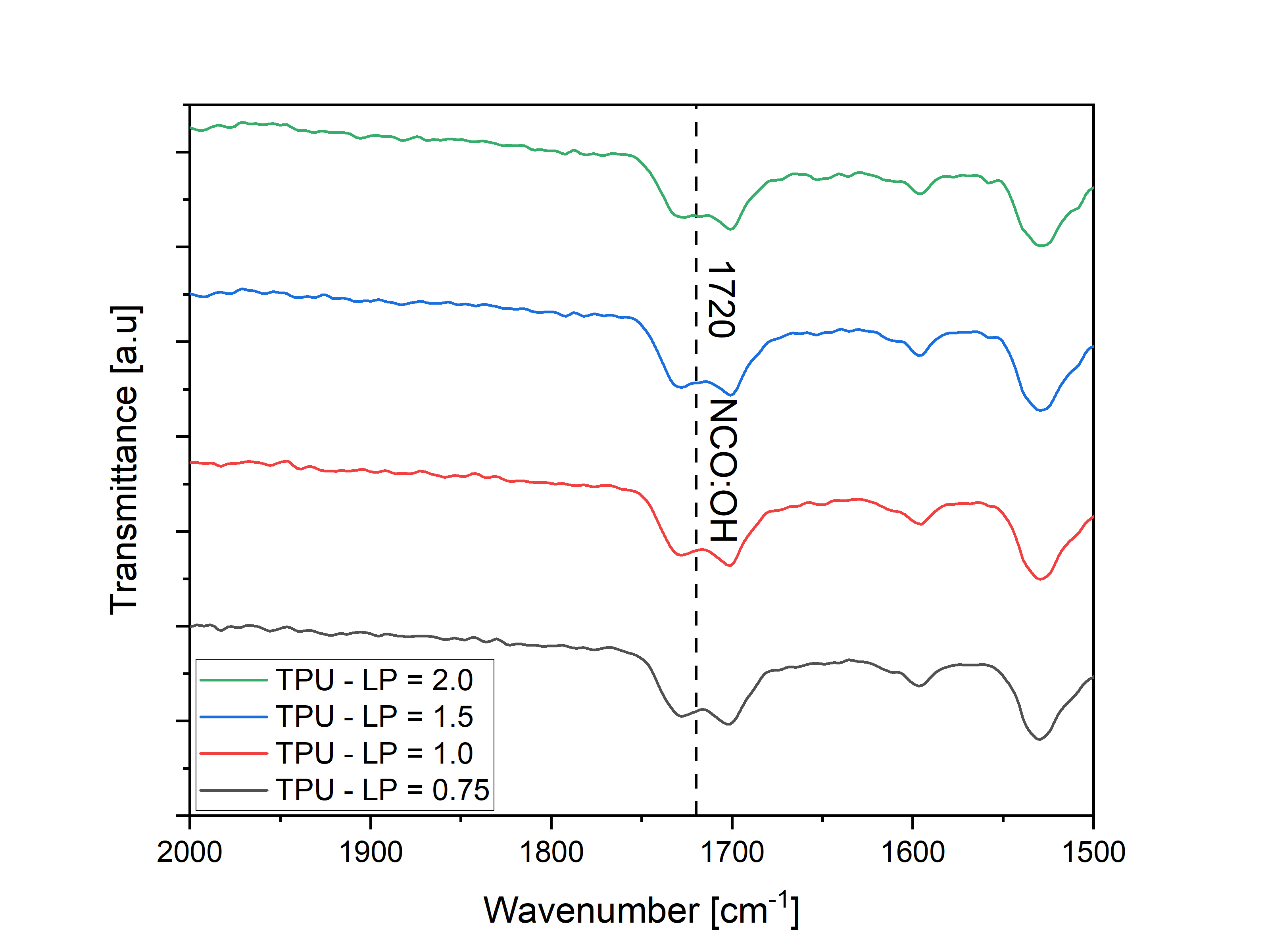}
}
\caption{FTIR curves for different laser powers (a) from 4000 $cm\textsuperscript{-1}$ to 400 $cm\textsuperscript{-1}$ and (b) from 2000 $cm\textsuperscript{-1}$ to 1500 $cm\textsuperscript{-1}$.}
\label{img:FTIRLP}
\end{figure}

\begin{figure}[h!]
\centering
\subfigure[]{
\includegraphics[width=0.45\textwidth]{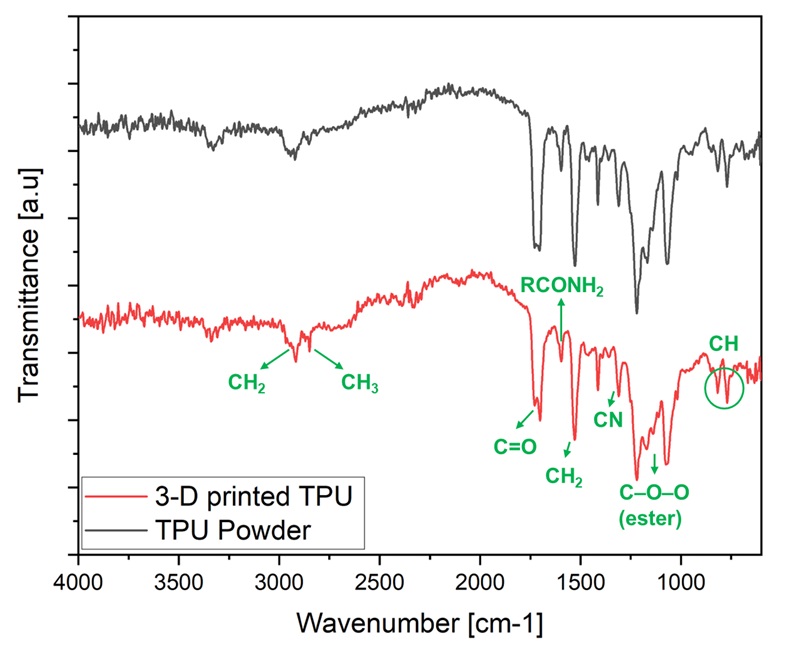}
}
\centering
\subfigure[]{
\includegraphics[width=0.45\textwidth]{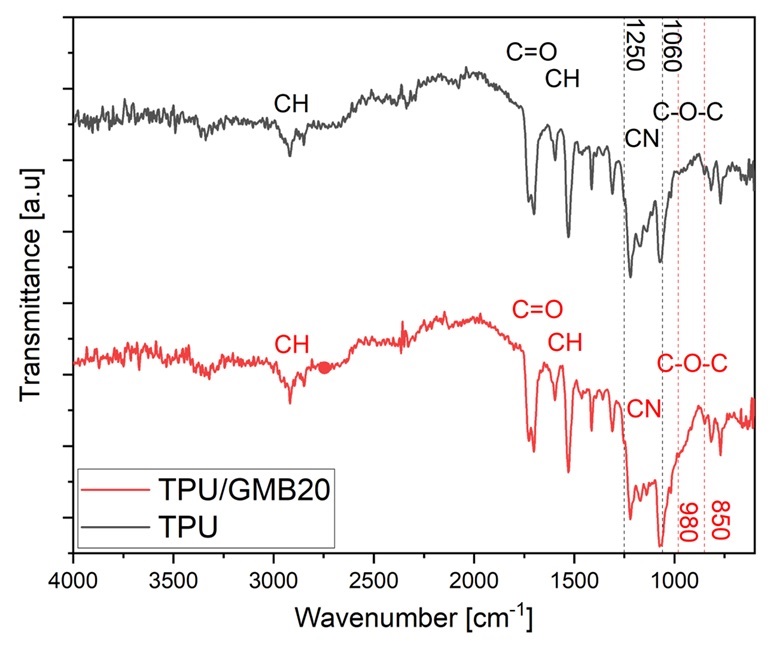}
}
\caption{FTIR curves for (a) TPU powder and the 3D printed pure TPU foam and (b) TPU foam and SF60-20 foam printed with same parameters.}
\label{img:FTIRMIX}
\end{figure}

We examined the chemical bonds within TPU and TPU/GMB foams using FTIR. In this study, we used a polyester-based TPU as can be seen from the FTIR spectroscopy graph shown in \textbf{Figure} \ref{img:FTIRMIX}, which shows the characteristic \chemfig{C=O} group in the polyurethane and NH stretching vibrations at 1741 $cm\textsuperscript{-1}$ and 1540 $cm\textsuperscript{-1}$, respectively \cite{Jin2020}. We also see two strong absorption peaks at 2960 $cm\textsuperscript{-1}$ and 2823 $cm\textsuperscript{-1}$ which is attributed to the stretch vibrations of \textbf{$CH_{2}$} and \textbf{$CH_{3}$}. When we increased the laser power ratio, effectively changing the supplied energy density to the powder, we observed that the absorption intensity of the hydrogen-bonded \chemfig{C=O} band (right of 1722 $cm\textsuperscript{-1}$) compared with the non-hydrogen bonded \chemfig{C=O} band (left of 1722 $cm\textsuperscript{-1}$) increased as observed in Figure~\ref{img:FTIRLP}. This behavior is a consequence of the formation of more organized hard segments in the TPU's micro-phase, resulting in a higher percentage of hard domains relative to soft domains.  

Next, comparing the FTIR spectra in Figure~\ref{img:FTIRMIX}(b) of pure TPU and TPU/GMB foams printed using SLS, we only observed new peaks close to 980 $cm\textsuperscript{-1}$ and 850 $cm\textsuperscript{-1}$, which correspond to the presence of \chemfig{Si-O-Si}. This shows that no new chemical bonds were formed between the TPU matrix and GMBs, and hence, the bonding was completely physical. \cite{zhang2018crack}. 
%From the measured FTIR spectra of GMBs (Figure \ref{img:FTIRMIX}), we observed sharp absorption peaks at 980 $cm\textsuperscript{-1}$ and 850 $cm\textsuperscript{-1}$, which corresponds to \chemfig{Si-O-Si} the borosilicate material in GMBs. When TPU and GMBs are combined to manufacture a 3D printed part, we performed FTIR to see if there were any chemical changes due to the addition of these GMBs. 

\subsubsection{Influence of print and foam parameters on micro-structure}\label{microstructure}

\begin{figure}[h!]
    \resizebox{1\textwidth}{!}{
    \begin{tabular}{|c|c|c|c|c|}
    \hline
  \textbf{LH / LPR}  & 0.75 & 1.0 & 1.5 & 2.0 \\
        \hline
\rotatebox{90}{75{$\mu{m}$}} & \includegraphics[width=0.30\textwidth]{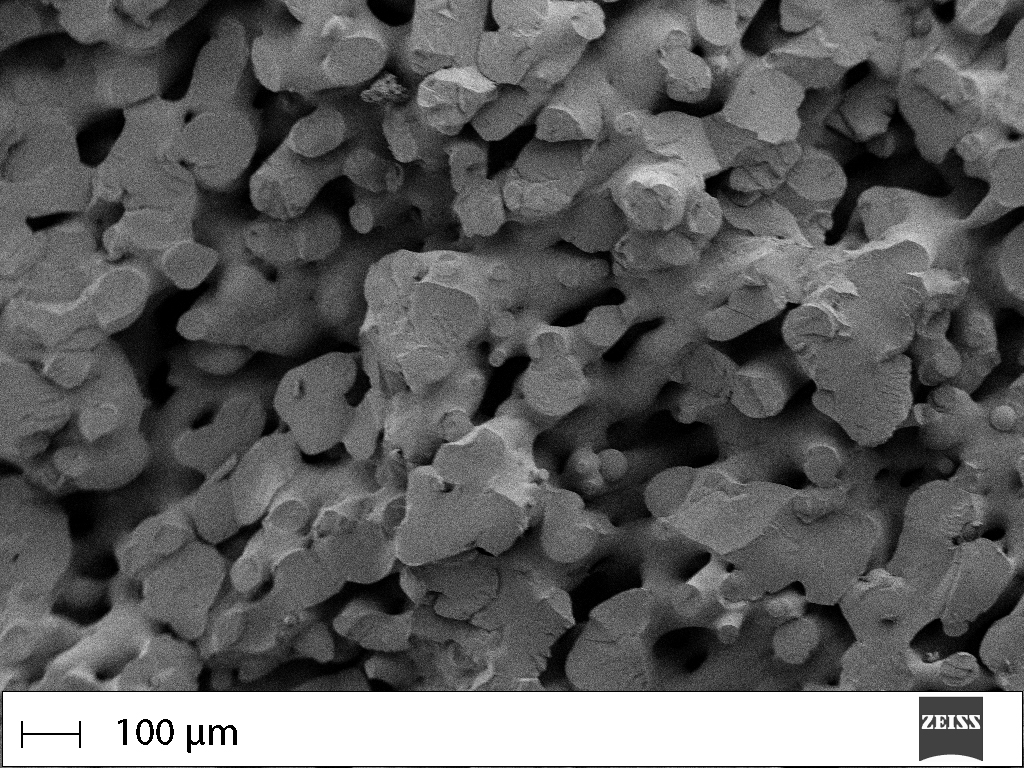} (a) &  \includegraphics[width=0.30\textwidth]{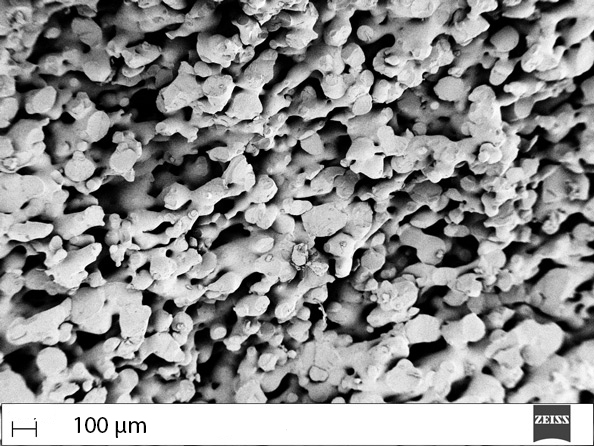} (b)
& \includegraphics[width=0.30\textwidth]{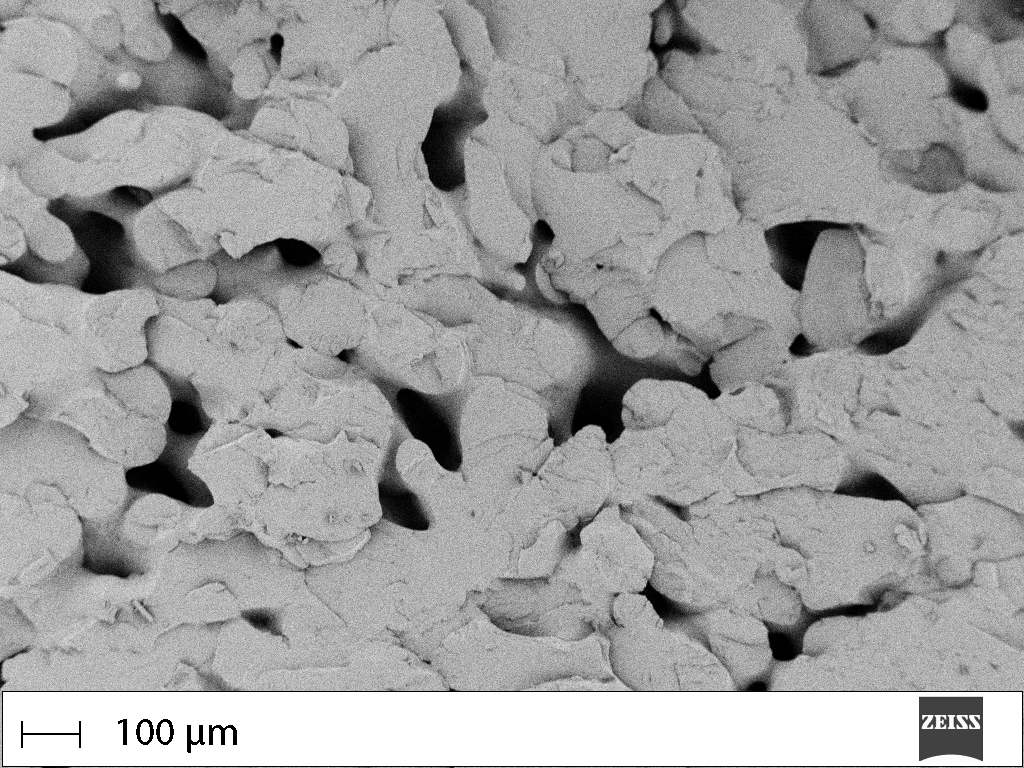} (c) & \includegraphics[width=0.30\textwidth]{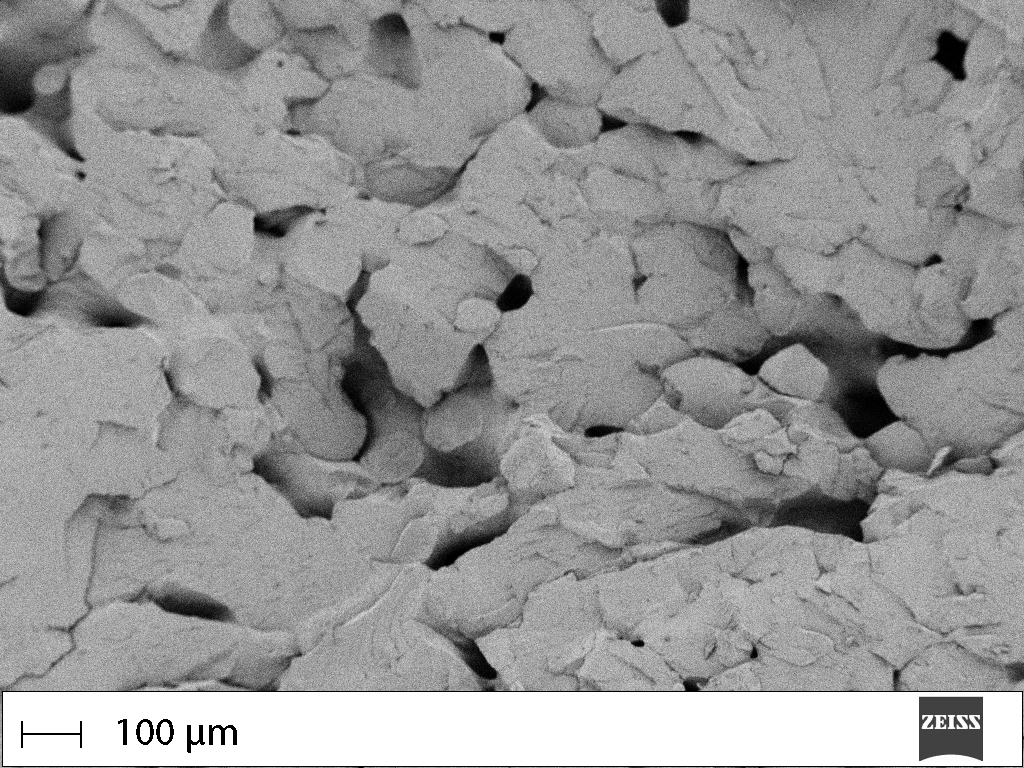} (d) \\   
        \hline
\rotatebox{90}{125{$\mu{m}$}} & \includegraphics[width=0.30\textwidth]{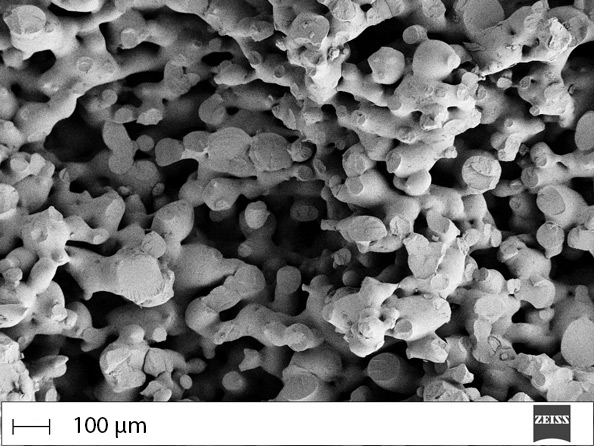} (e) &  \includegraphics[width=0.30\textwidth]{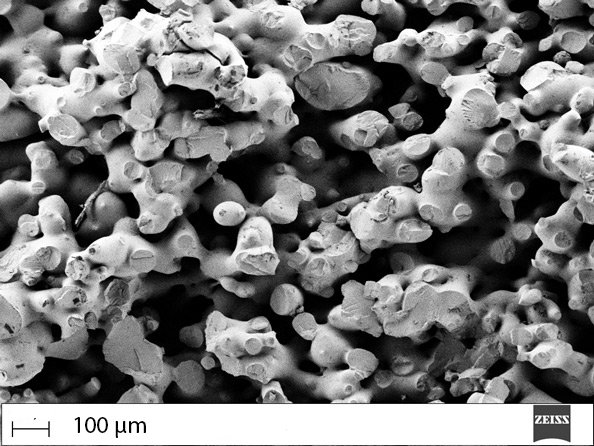} (f)
& \includegraphics[width=0.30\textwidth]{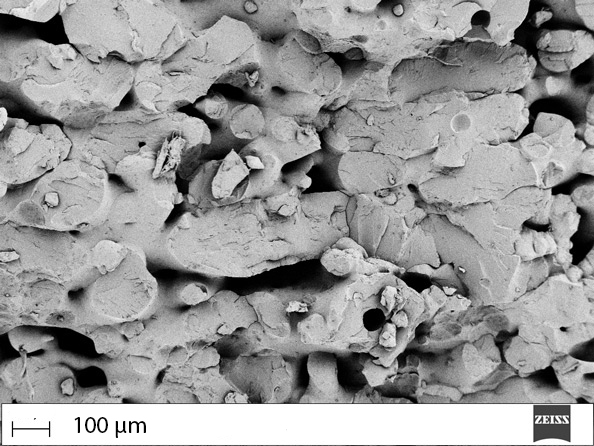} (g) & \includegraphics[width=0.30\textwidth]{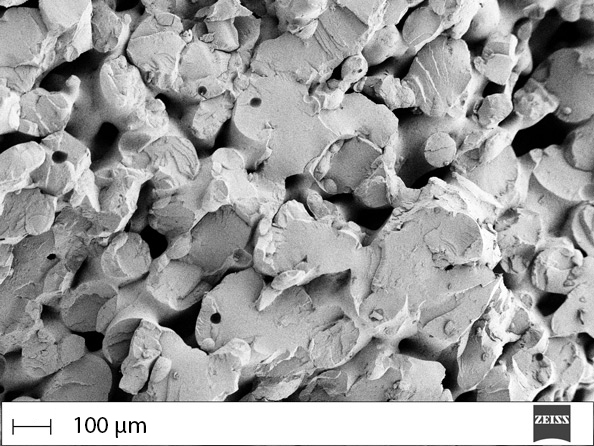} (h)\\   
        \hline
        
                \hline
\rotatebox{90}{175{$\mu{m}$}} & \includegraphics[width=0.30\textwidth]{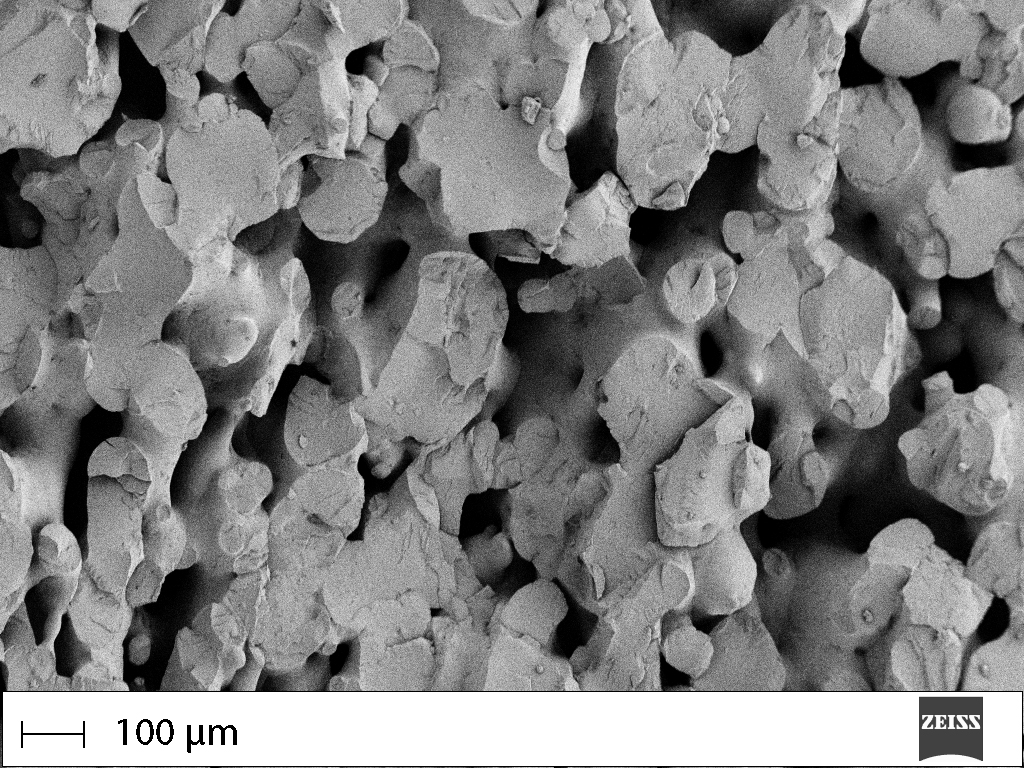} (i) &  \includegraphics[width=0.30\textwidth]{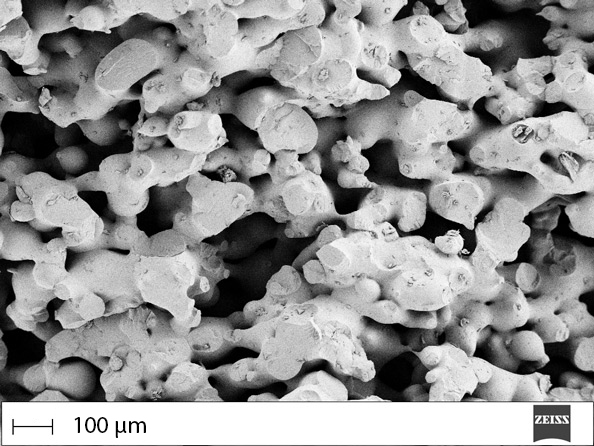} (j)
& \includegraphics[width=0.30\textwidth]{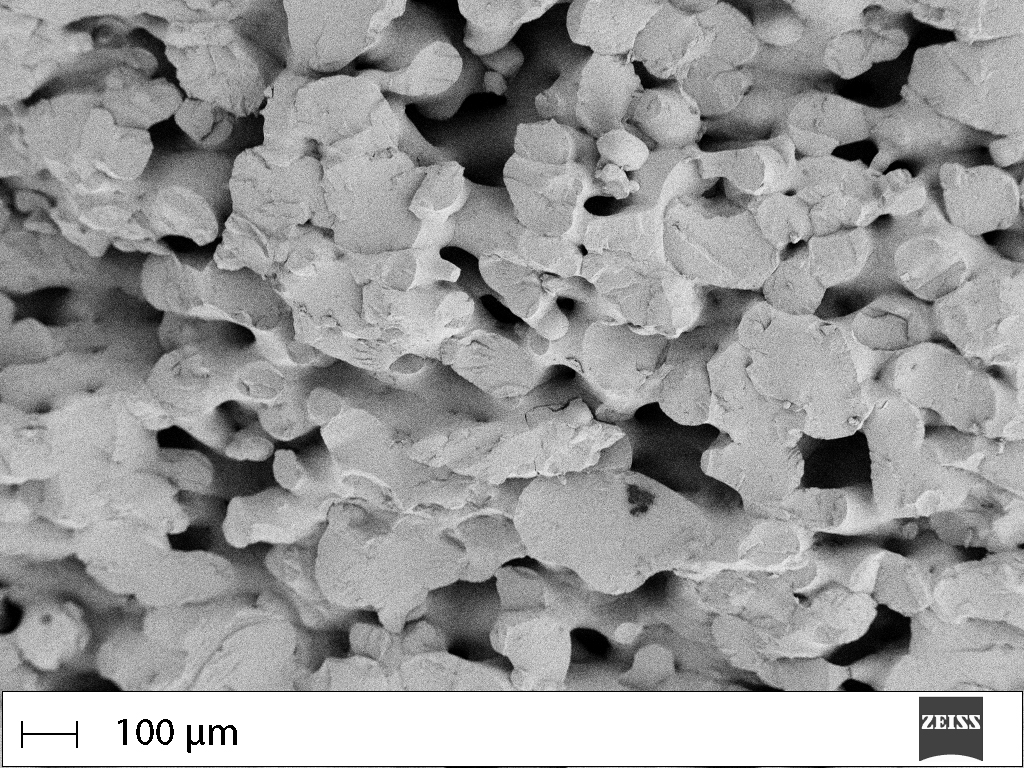} (k) & \includegraphics[width=0.30\textwidth]{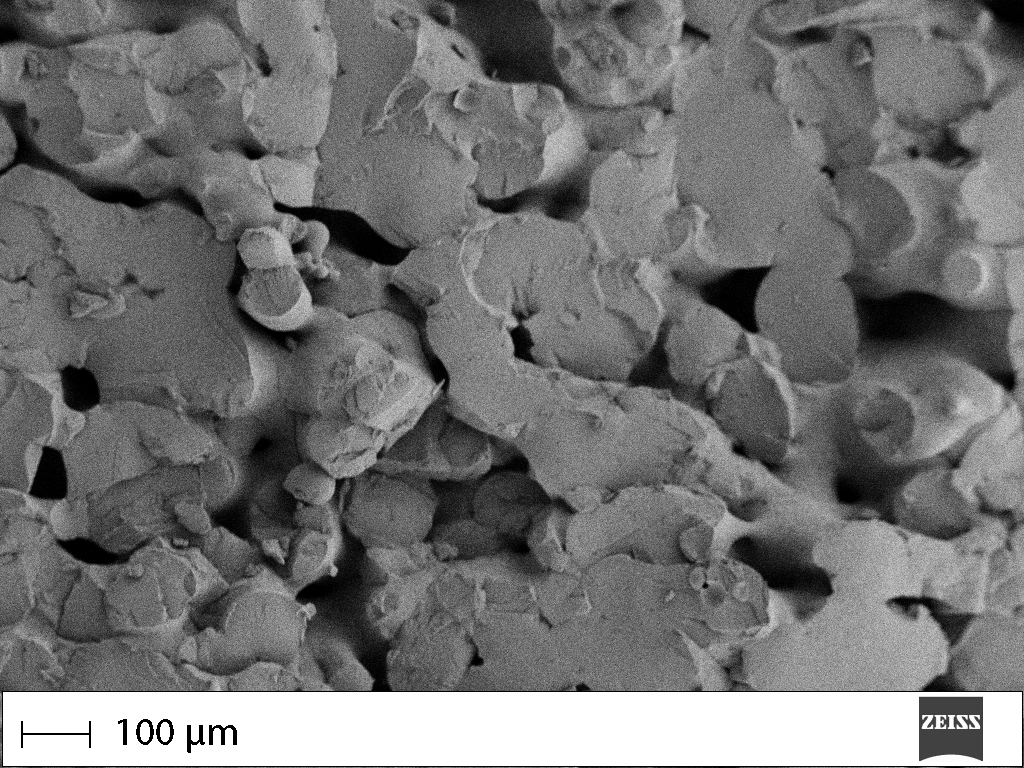} (l) \\   
        \hline
        
    \end{tabular} }
        \caption{SEM images (E.H.T. = 3kV and Signal = SE2) of morphologies observed for pure TPU foams printed with different Laser Power Ratio (LPR) and Layer Height (LH) combinations.}
\label{img:SEM-printP}
\end{figure}

Pure TPU foams manufactured using the print parameters specified in \textit{Section} \ref{EffPP} were analyzed under SEM to observe the effect of these parameters on the microstructure of these foams. From \textbf{Figure} \ref{img:SEM-printP}, we see that the TPU foams manifested a controlled porous structure with a segregated matrix as observed in previous studies \cite{li2017selective,Yuan2017,Toncheva2021}. In this study, the wall size of the segregated matrix at the microscale is referred to as \textit{cell wall thickness}. When we increased the laser power ratio (LPR) from 0.75 to 2.0, the cell wall size appeared to increase, manifesting denser foams. This is attributed to an increase in the melt pool size due to higher energy supplied. We did not see a distinctive change in foam morphology when layer height decreased for LPR values of 0.75 and 1.0. However, for higher LPR values of 1.5 and 2.0, we observed that the cell wall thickness increased with decreasing layer height. This is attributed to the repeated sintering of more layers due to smaller layer heights. The implications of LPR values and layer heights on the mechanical response will be further elaborated in \textit{Section} \ref{TensRes}.

\begin{figure}[h!]
    \resizebox{1\textwidth}{!}{
    \begin{tabular}{|c|c|c|c|}
    \hline
  \textbf{Print Parameters / Foams}  & SF60-20 & SF22-20 & SF15-20  \\
        \hline
\rotatebox{90}{LH = 75{$\mu{m}$} , LPR = 1.5} & \includegraphics[width=0.30\textwidth]{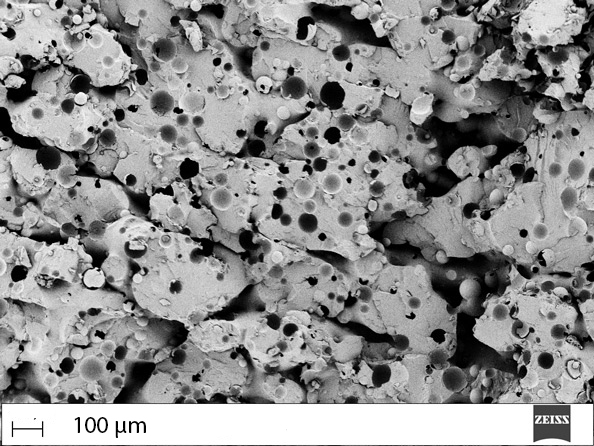} (a) &  \includegraphics[width=0.30\textwidth]{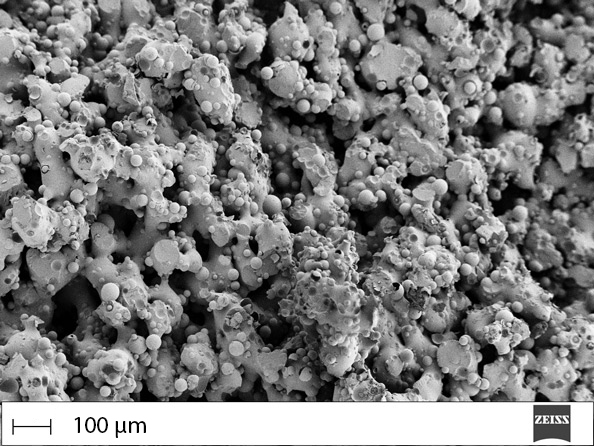} (b)
& \includegraphics[width=0.30\textwidth]{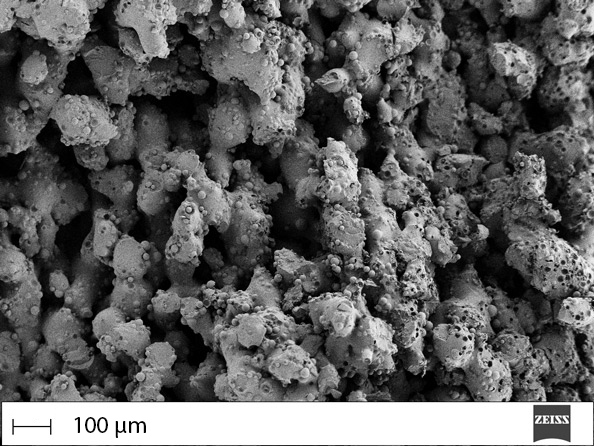} (c) \\   
        \hline
\rotatebox{90}{LH = 75{$\mu{m}$} , LPR = 2.0} & \includegraphics[width=0.30\textwidth]{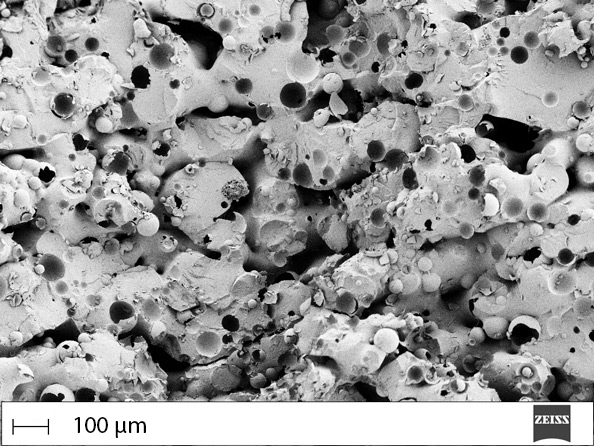} (d) &  \includegraphics[width=0.30\textwidth]{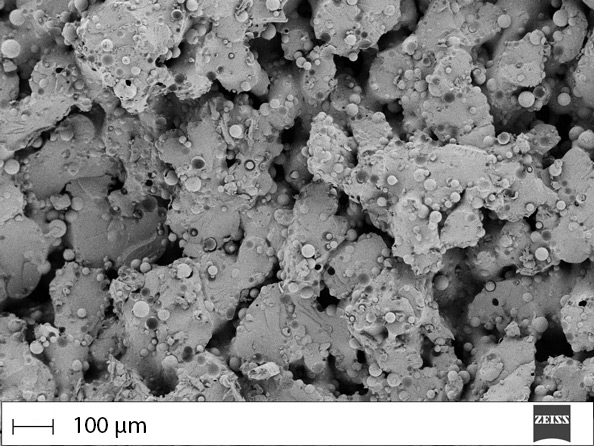} (e)
& \includegraphics[width=0.30\textwidth]{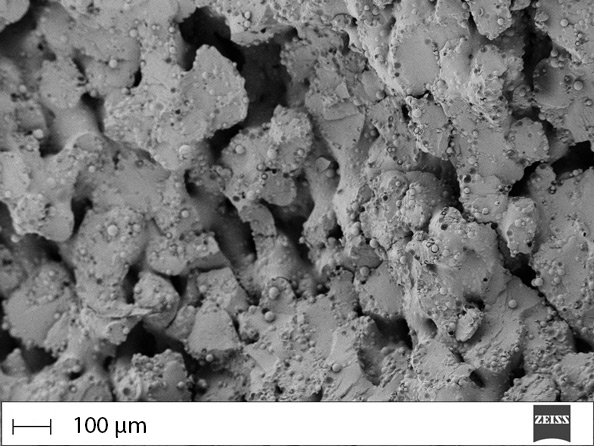} (f) \\   
        \hline
        
    \end{tabular} }
        \caption{SEM images (E.H.T. = 3kV and Signal = SE2) of foams printed with final print parameters (Laser Power Ratio = 1.5 and Layer Height = 75 {$\mu{m}$}) containing 20\% volume fraction of GMBs for (a) SF60-20; (b) SF22-20; and (c) SF15-20; and 
SEM images of foams printed with the modified print parameters (Laser Power Ratio = 2.0 and Layer Height = 75 {$\mu{m}$}) containing 20\% volume fraction of GMBs for (d) SF60-20; (e) SF22-20; and (f) SF15-20}
\label{img:SEM_diffP}
\end{figure}

To evaluate the effect of the size of the GMBs, we printed foams with the final parameters chosen based on the mechanical performance, details of which are provided later in \textit{Section} \ref{TensRes}. \textbf{Figure} \ref{img:SEM_diffP} shows the SEM images of syntactic foams printed with different size GMBs (SF60 - large, SF22 \& SF15 - small) at a fixed volume fraction of 20\%. We observed that bigger GMBs tend to lodge between cell walls and in the cell walls in SF60-20. However, in SF22-20 and SF15-20 with smaller GMBs, the GMBs tend to get lodged within the cell walls of the TPU foam due to the larger space between the cell walls than the particle size. The cell walls of the SF22-20 and SF15-20 foams are also thinner than those of the SF60-20 foams. As explained above, this drop in cell wall size is related to the decrease in energy absorbed by the TPU powder due to the presence of GMBs with a high particle per area density. In this study, the energy delivered to all polymer blends was held constant to correlate the GMB parameters with the print parameters, and further evaluate the mechanical performance of the printed syntactic foams. However, a larger value of supplied energy density may be selected for smaller GMBs to achieve the same cell wall thickness for all blends. To show the effect of energy density on the morphology (cell wall thickness) of the TPU/GMB syntactic foams, we printed SF60-20, SF22-20, and SF15-20 foams with an LPR of 1.5 and 2.0 and a layer height of 75 {$\mu{m}$}. We see in Figure~\ref{img:SEM_diffP} (d)-(f) that the cell walls for the smaller particles appear thicker compared to foams shown in \ref{img:SEM_diffP} (a)-(c) as we increased the LPR. Therefore, the energy supplied to the system must be increased for the TPU/GMB powder blends to avoid reducing the wall thickness.

\subsubsection{Porosity}\label{results:porosity}
We examined the porosity of TPU-based syntactic foams to determine the effect of introducing GMBs into the matrix during additive manufacturing. This porosity is the void content between the cell walls of the TPU, and does not consider the voids that exist within the GMB particles. The porosity values for TPU, SF60-20, and SF60-40 foams remained constant, ranging between 27 and 28 percent. However, the SF60-60 foam porosity increased significantly to 38 percent, which can be attributed to an insufficient matrix available to bind together a high volume fraction of GMBs. By altering the energy density supplied to the system, the porosity values of GMB-containing foams can decrease as was observed with the increase in cell wall thickness for pure TPU in the \textit{section} \ref{microstructure}. From Figure \ref{img:SEM_diffP}, we also noticed that for the same volume fraction of GMBs, decreasing the size of the GMBs decreased the cell wall size, hence increasing the porosity of the foam. An increase in particle density of the TPU/ GMB blend altered the transmittance and absorbance of the mixture (\textit{Section} \ref{PSD}). This suggests that raising the energy density of the laser with increasing GMB content can decrease the porosity as can be observed in Figure \ref{img:SEM_diffP}, comparing (a) - (d), (b) - (e), and (c) - (f). The same porosity can be achieved if higher energy density is supplied to a powder blend with smaller particles as compared to low energy density supplied to a blend without particles - this phenomenon can also be observed in Figure \ref{img:SEM_diffP} comparing (a) - (e).

\subsection{Mechanical Response} \label{MechRes}

In this section, we first discuss the effect of print parameters on the tensile properties of pure TPU foams. Then, the performance of syntactic foams printed with the chosen print parameters under tensile and compression loading with various volume fractions and GMB types is evaluated. In addition, failure morphologies are discussed to complement the response.

\subsubsection{Tensile Response} \label{TensRes}
\paragraph{{\bf\em Pure TPU foams}}

\begin{figure}[H]
\centering
\subfigure[]{
\includegraphics[width=0.45\textwidth]{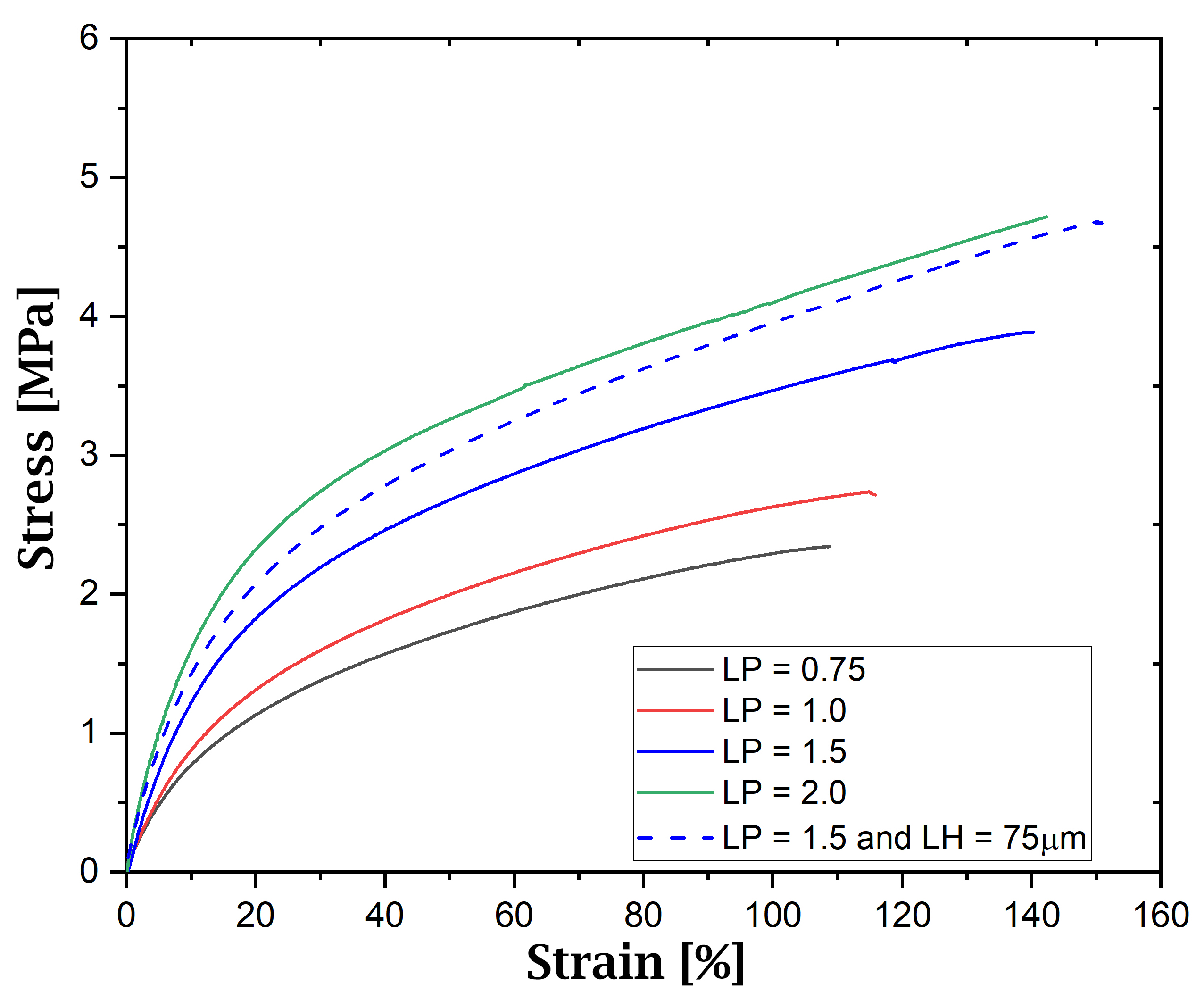}
\label{img:TensLPLH1}
}\hspace{3mm}
\centering
\subfigure[]{
\includegraphics[width=0.48\textwidth]{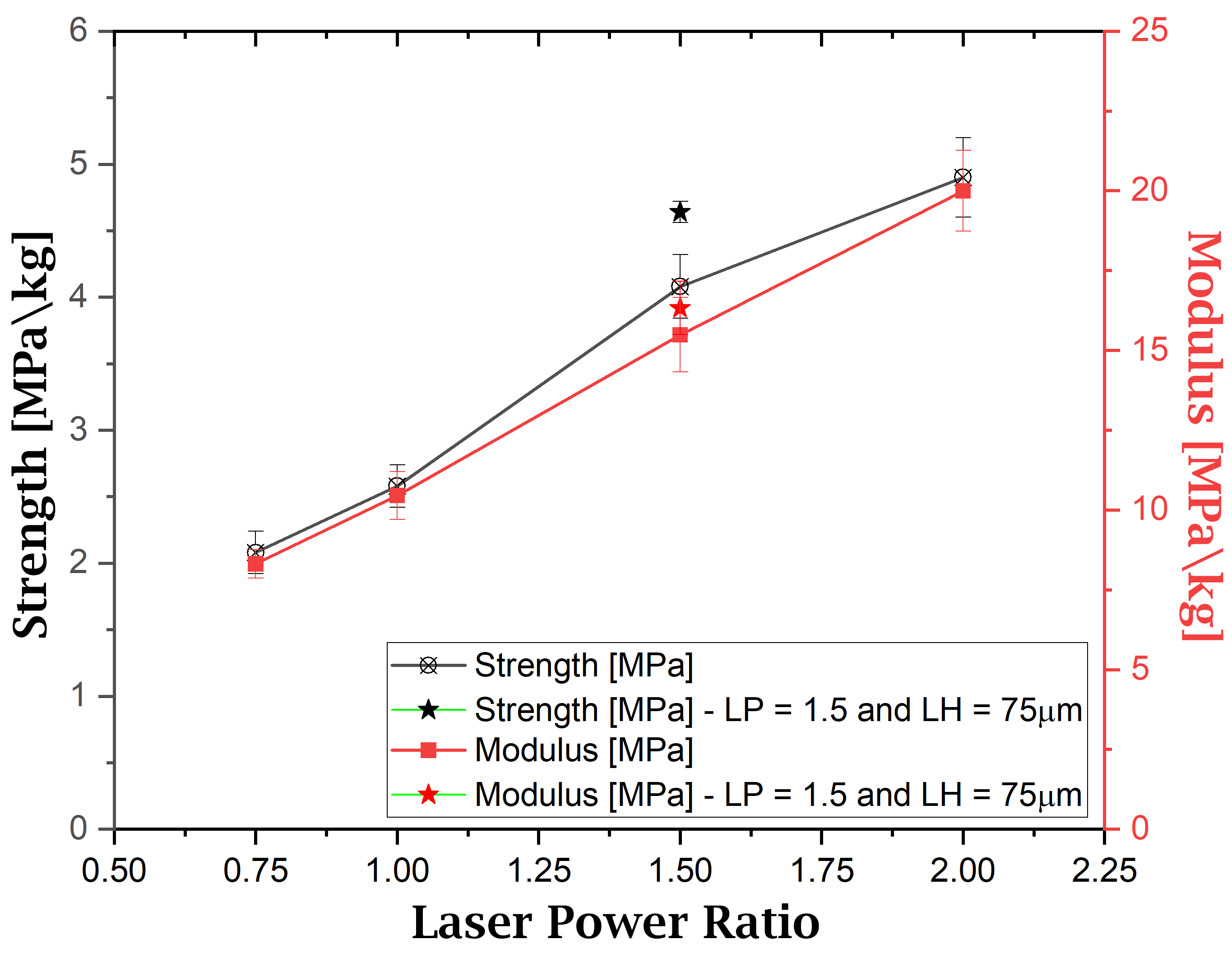}
\label{img:TensLPLH2}
}
\caption{(a) Representative tensile stress-strain plots for pure TPU foams with changing laser power ratio and a fixed layer height (layer height was varied for a selected laser power); (b) Summary of tensile properties with varying print parameters.}

\end{figure}

\textbf{Figure} \ref{img:TensLPLH1} shows representative stress-strain curves for all 3D printed pure TPU foams, and \textbf{Figure} \ref{img:TensLPLH2} summarizes tensile properties (modulus and strength) from all the tests performed. These foams exhibited a small linear elastic region followed by a nonlinear ductile response under tension. By increasing the laser power ratio (LPR) at a constant layer height (LH) value of 125 {$\mu m$}, we observed that the tensile modulus and strength increased. This is attributed to an increase in cell wall thickness and reduced porosity with an increase in LPR, as observed in the SEM images in Section \ref{microstructure}. The strength increased because of an increase in the load-bearing phase resulting in better load distribution. Moreover, an increase in the cell wall size increased the strain to failure. However, with an increase in the energy supplied value at a higher LPR, we also observed that the dimensional stability decreased, that is, we observed higher z-bulging. Therefore, to avoid this, we printed a specimen at an LPR of 1.5 and a reduced LH of 75 {$\mu m$}. When we decreased the LH, tensile strength increased drastically by 10.87 \% compared to LH of 125 {$\mu m$} shown by black star marker in \textbf{Figure} \ref{img:TensLPLH2} , while tensile modulus increased moderately by 0.62 \%. This is attributed to the repeated sintering of layers as we decreased the LH, which increased the energy supplied to individual layers and also, provided better bonding even at a lower LPR of 1.5.

The influence of the two parameters, LPR and LH, was considered in unison to determine optimal final print parameters for enhanced tensile response and high dimensional stability of the printed specimens. Considering the tensile performance, print duration, and printer limitations, we chose an LPR of 1.5 and an LH of 75 {$\mu m$} for the next printing steps.

\paragraph{{\bf\em Particle reinforced TPU foams}}

\begin{figure}[H]
\centering
\subfigure[]{
\includegraphics[width=0.45\textwidth]{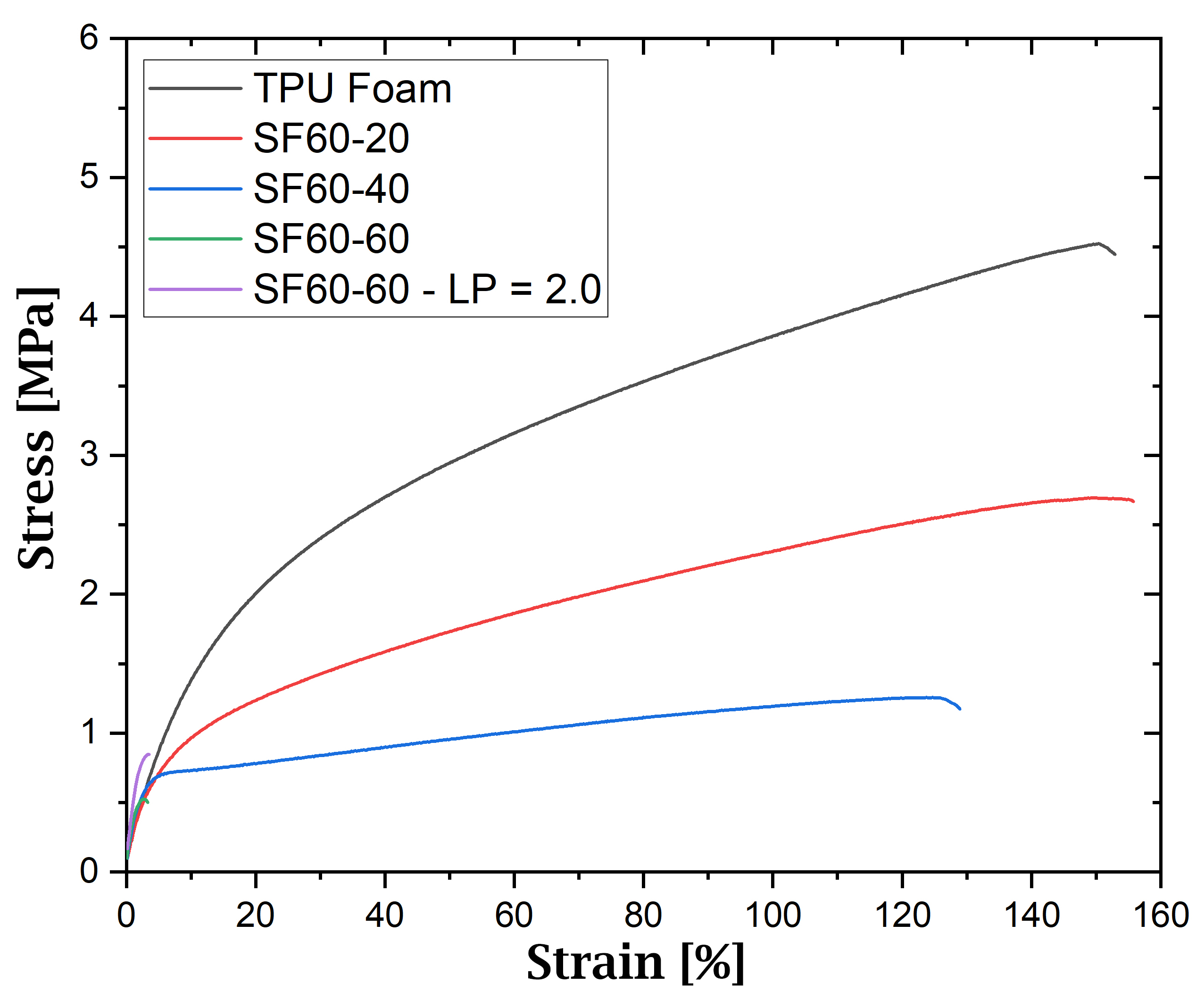}
}\hspace{3mm}
\centering
\subfigure[]{
\includegraphics[width=0.45\textwidth]{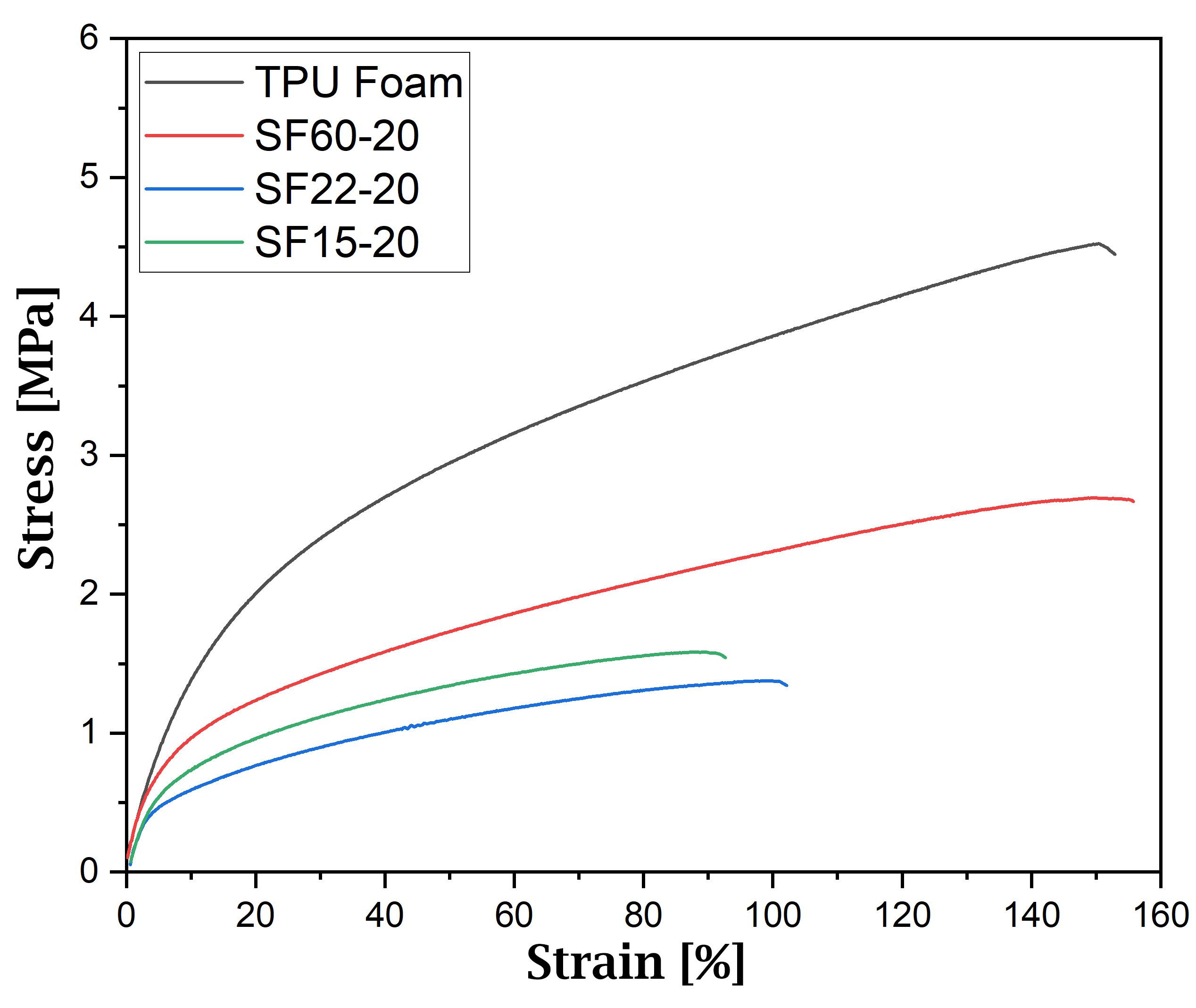}
}
\caption{(a) Representative tensile stress-strain plots for syntactic foams with changing GMB volume fraction (for volume fraction 60\%, a laser power ratio of 2.0 was also used); (b) Representative tensile stress-strain plots for syntactic foams with varying grades of GMBs with fixed volume fraction = 20\%.}
\label{img:Tens_SS}
\end{figure}

Using the final print parameters established for pure TPU foams, we printed TPU-based syntactic foams and tested them under quasi-static tensile loading. We first fixed the GMB size to larger particles and varied their volume fraction, followed by fixing the volume fraction at 20\% and varying the GMB sizes. \textbf{Figure} \ref{img:Tens_SS}(a) shows representative tensile stress-strain responses for SF60 foams with varying GMB volume fractions. We see that the strength and strain to failure decreased with increasing GMB volume fraction of the larger particles. This behavior can also be attributed to poor adhesion of the larger particles partially embedded within a segregated matrix resulting in debonding and insufficient load transfer. SF60-60 with 60\% GMB volume fraction showed significantly poorer tensile strength and failure to strain properties compared to SF60-20 and SF60-40 due to the lack of TPU matrix to effectively bind the GMBs at very high volume fractions. On the other hand, the elastic moduli were in the range of 13 to 16 MPa for TPU, SF60-20, and SF60-40 foams. We speculate that these small deviations are due to competing effects of stiffer GMB particles and reducing cell wall size with increasing GMB volume fraction. Even though the GMBs are stiffer than TPU, their effect is counteracted by a reduction in the cell wall thickness when printed with the same supplied laser energy density.

Figure \ref{img:Tens_SS}(b) shows the representative tensile stress-strain responses of syntactic foams with three GMB particle sizes at 20\% GMB volume fraction. We observed that the tensile strength and strain to failure reduced when we incorporated smaller GMBs (GM22 and GM15) than larger GMBs (GM60) at the same volume fraction, rendering the foam quasi-brittle. This is attributed to two aspects: 1) Smaller GMBs tend to get lodged in the cell walls, resulting in more stress concentration locations within the cell wall, and 2) Smaller particles have higher particle density than larger particles for a fixed foam volume. This results in a higher surface area of particles in contact with the matrix causing higher debonding cites compared to that of the SF22-20 foams, thereby, reducing the tensile strength. Tensile moduli values of SF22-20 and SF15-20 dropped to an average of 12.8 and 11.6 MPa compared to 16.3 MPa for the pure TPU foam. We attribute this reduction in moduli to the cell wall thickness reduction associated with higher particle density in SF22-20 and SF15-20 foams compared to that of SF60-20 as shown previously in Figure~\ref{img:SEM_diffP}.    

Tensile properties for different grades of GMBs at different volume fractions are summarized in \textbf{Figure} \ref{img:Comp_PlotNorm}.

\subsubsection{Compressive Response}
We performed uniaxial compression tests to determine the mechanical response of pure and particle-reinforced TPU foams. We first selected GM60 particles and varied the volume fractions ranging from 20\% to 60\% in increments of 20\% and incorporated them in the segregated TPU matrix. Further, to understand the effect of particle size within the segregated structure of the matrix, we choose GM22 and GM15 GMBs with volume fractions of 20\% and 40\%. In this paper, the compressive strength is chosen as the compressive stress value at 30\% strain value.

\begin{figure}[h!]
\centering
%\subfigure[]{
\includegraphics[width=0.8\textwidth]{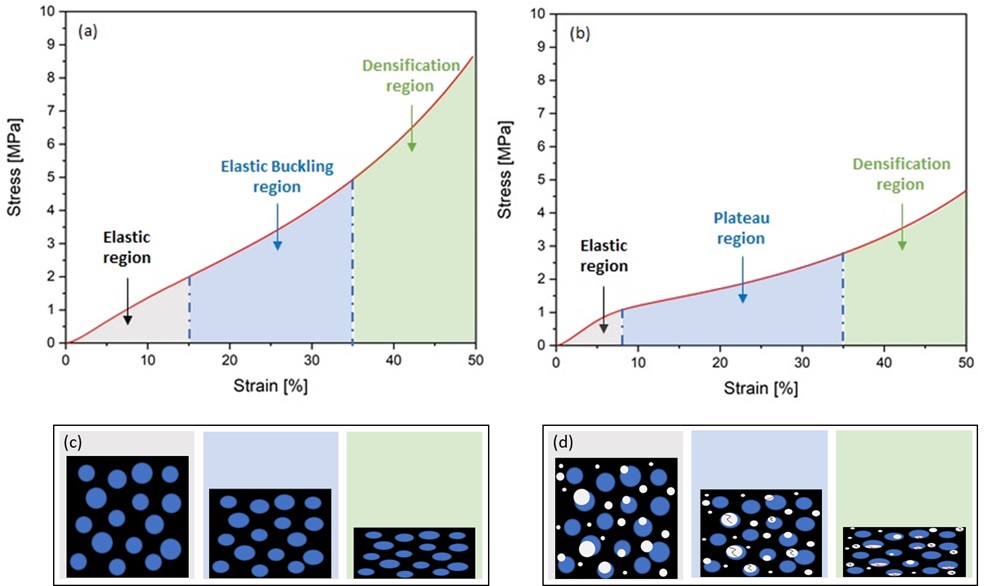}
%}
\caption{Typical compressive stress-strain behavior of (a) pure TPU foams and (b) syntactic foams, and Illustrations of the compression mechanics for (c) pure TPU foams and (d) syntactic foams.}
\label{img:CompAnim}
\end{figure}

\begin{figure}[h!]
\centering
\subfigure[]{
\includegraphics[width=0.45\textwidth]{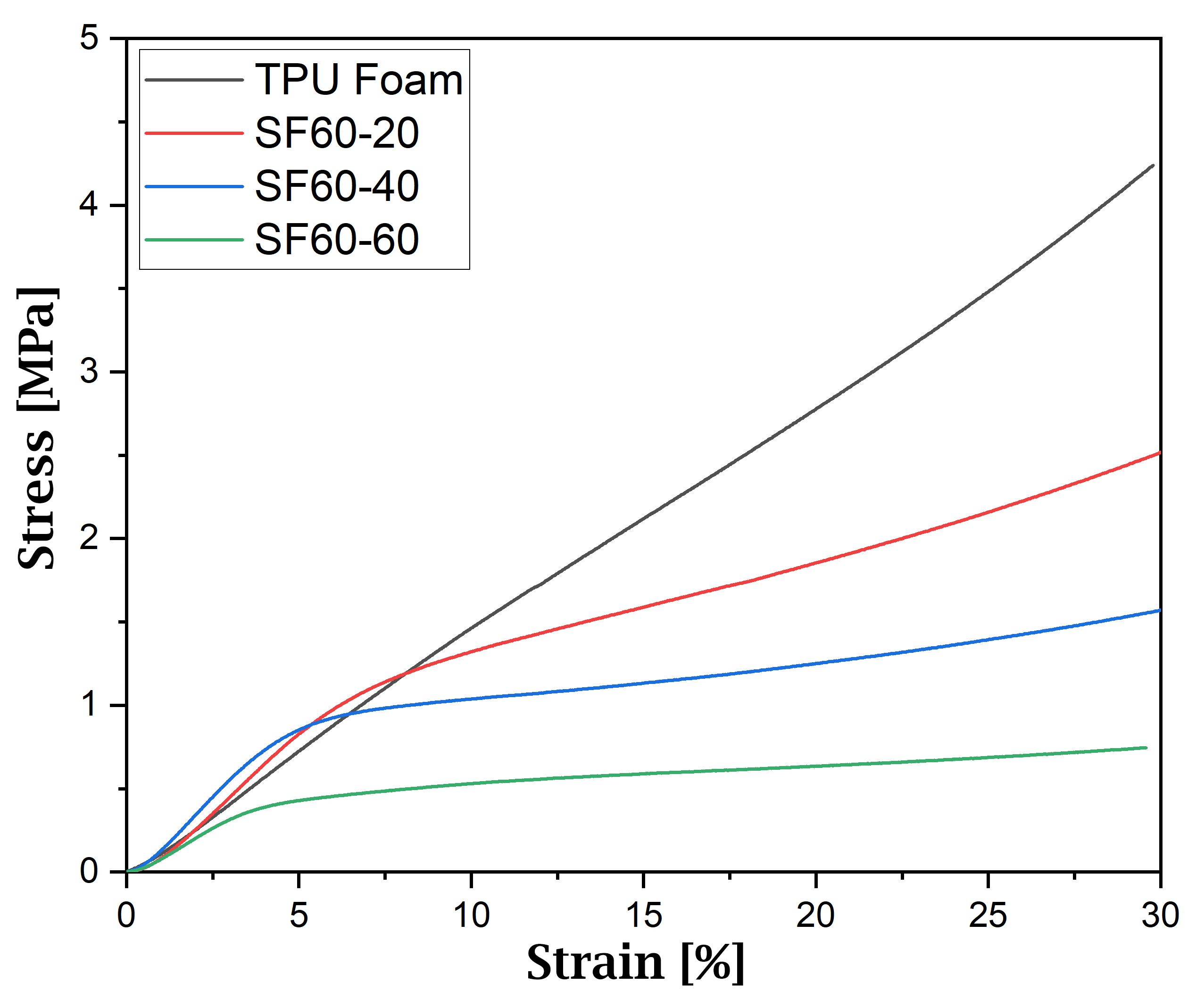}
}\hspace{3mm}
\centering
\subfigure[]{
\includegraphics[width=0.45\textwidth]{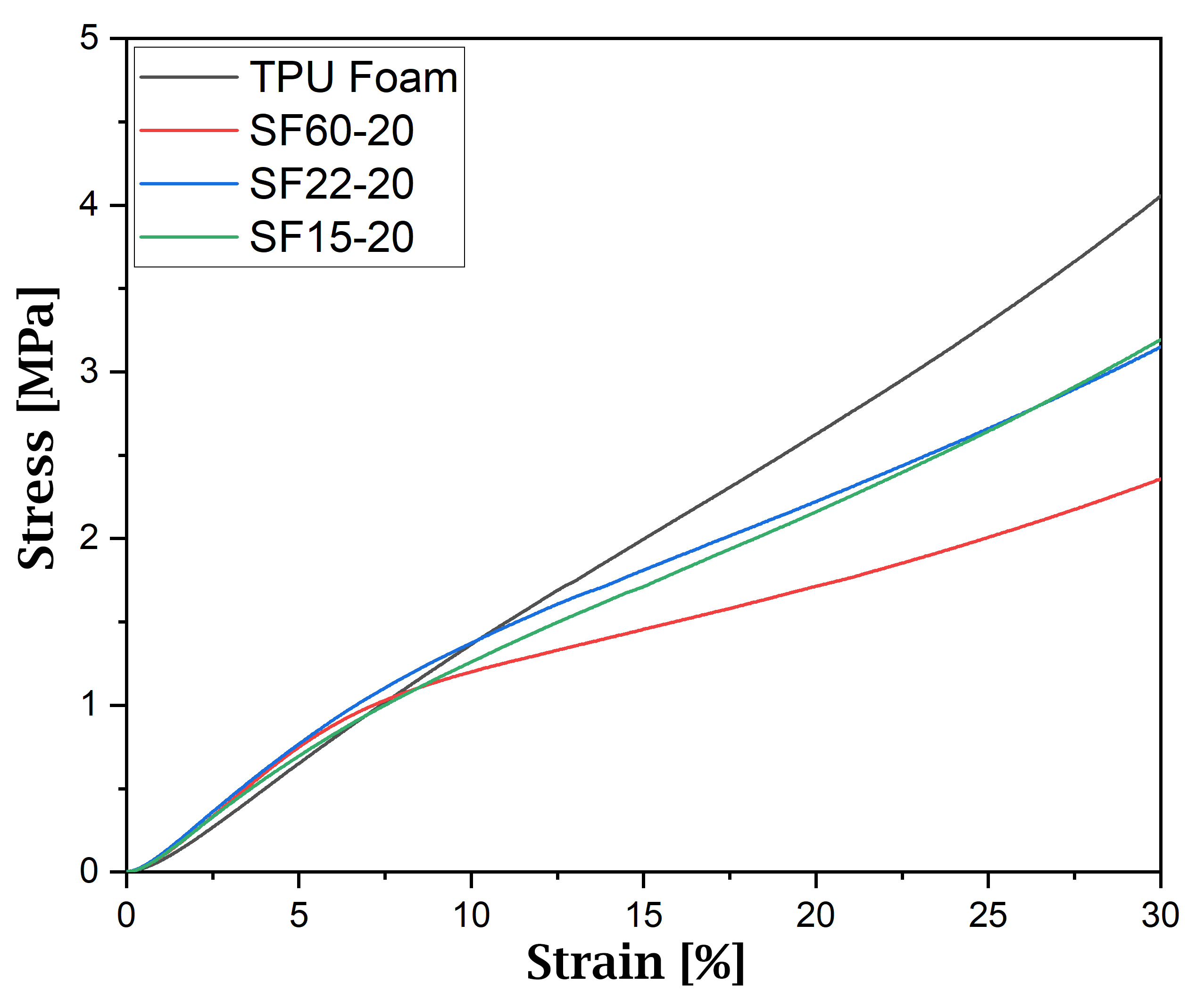}
}
\caption{(a) Representative compressive stress-strain plots for SF60 foams with changing GMB volume fraction; (b) Representative compressive stress strain plots for syntactic foams with varying grades of GMBs with fixed volume fraction = 20\%.}
\label{img:Comp_Plot}
\end{figure}

The stress-strain response for pure TPU foams resembled the non-linear behavior of typical foamed elastomers \cite{roman2022} as shown in \textbf{Figure} \ref{img:CompAnim}(a). Due to the segregated matrix micro-structure (shown as blue regions) in these foams, we observed a small elastic region at the beginning which is then followed by the elastic buckling zone where the cell walls start to buckle. It should be noted that the matrix segregation is intentional, and it gives rise to a controlled porous structure. As the air is pushed out of the foams, the cell walls compress on themselves resulting in the densification zone. \textbf{Figure} \ref{img:Comp_Plot} shows a representative compressive stress-strain response of printed TPU foam. A schematic of a typical compressive stress-strain behavior for syntactic foams (Figure \ref{img:CompAnim}(b)) consists of a very small initial linear region associated with the enhanced elastic modulus due to the presence of reinforcing particles. This was followed by a second zone called the plateau region, corresponding to reduced stiffness due to particle cell wall buckling and crushing. Finally, the third zone, known as the densification region, occurs when the cell walls of the matrix and the particles compress.

\begin{figure}[H]
\centering
\subfigure[]{
\includegraphics[width=0.45\textwidth]{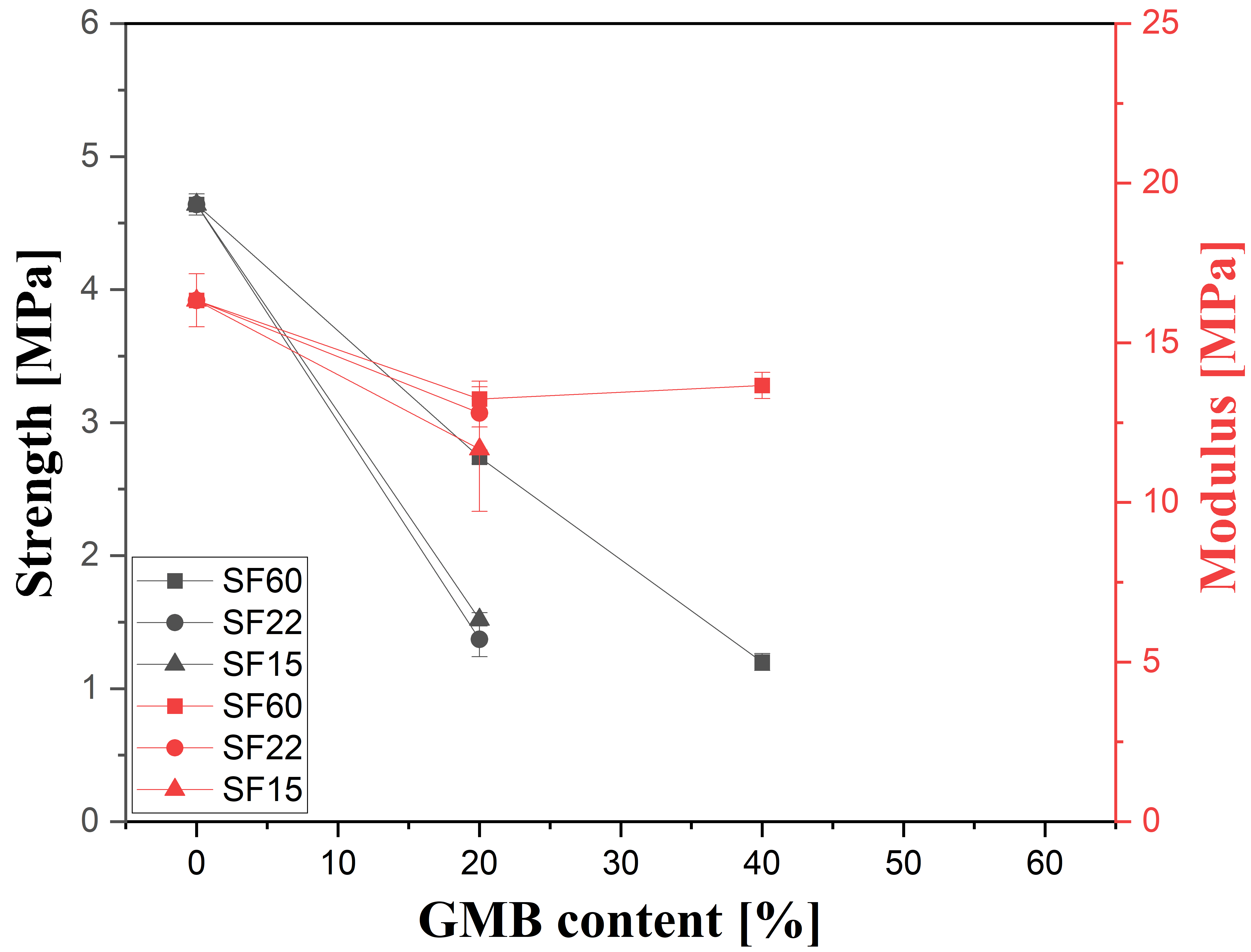}
}\hspace{3mm}
\centering
\subfigure[]{
\includegraphics[width=0.45\textwidth]{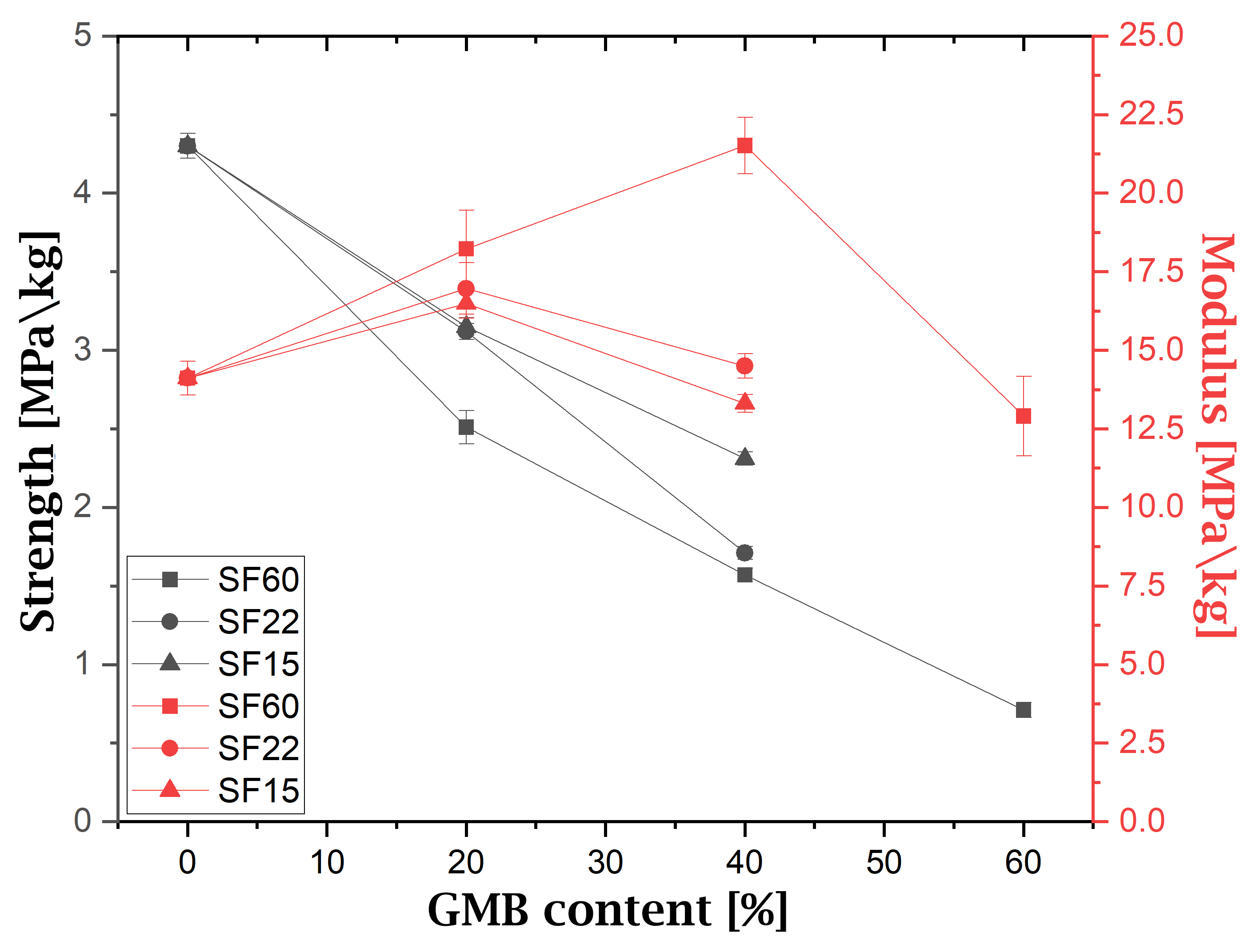}
}
\caption{Summary of (a) tensile properties and (b) compressive properties, for different GMB types with varying GMB content (volume fraction).}
\label{img:Comp_PlotNorm}
\end{figure}

\subparagraph{\textit{Different GMB Volume Fractions:}}\label{Comp_diffVF}
\textbf{Figure} \ref{img:Comp_Plot}(a) shows the representative compressive stress-strain responses of SF60 foams with varying GMB volume fractions. In general, we see that these responses exhibited an initial elastic region associated with the enhanced compressive modulus compared to the pure TPU foams in this region. This is attributed to the large particles lodged within and between the cell walls resulting in cell wall stiffening as well as creating stiff bridges between the cell walls of the segregated matrix. However, we noticed that the modulus reduced in SF60-60 foams due to insufficient matrix material available to bond these GMBs for effective load transfer to occur. At strain values beyond this initial region, we observed a knee formation due to the initiation of particle crushing in the gaps of the segregated matrix. A dominant plateau region after this knee formation is a characteristic behavior of particle crushing as discussed above in reference to Figure \ref{img:CompAnim}(b). We observed that the densification region is lower in the syntactic foams than that of the pure TPU foams, and it reduced with increasing GMB volume fraction. This is because of the lower crushing strength of GM60 particles. The compressive modulus and strength for SF60 foams are summarized in \textbf{Figure} \ref{img:Comp_PlotNorm}(b) using square markers.

\subparagraph{\textit{Different GMB Grades:}} \label{DiffGrade}

%For foams reinforced with smaller particles, as particles got lodged in the walls, there was no plateau region observed for these foams, the compression response was similar to that of the pure TPU foams. This reflects that the response for syntactic foams with smaller particles was matrix-dominated, and the particle strength would not affect the behavior of these foams.

Figure \ref{img:Comp_Plot}(b) illustrates compressive stress-strain responses of syntactic foams with GM60, GM22, and GM15 GMB size particles at 20\% volume fraction. With decreasing GMB size (GM22 and GM15), compression behavior resembled that of pure TPU foam, and the plateau region disappeared compared to that observed for TPU foam with bigger particles (GM60). We hypothesize that this response reflects matrix material dominance and does not involve particle crushing. Since the smaller particles are entirely embedded within flexible TPU matrix cell walls, it will take larger strains to reach their high crushing strengths. Although GM15 has a higher crushing strength than GM22, SF15 and SF22 foams display similar compression behavior as smaller particles embedded in the TPU matrix never achieved sufficient stresses to crush them. Figure \ref{img:Comp_Plot}(b) shows that SF22 and SF15 foams have lower densification stress than pure TPU foam. As mentioned previously in \textit{section} \ref{results:porosity}, printing syntactic foams with smaller GMBs increased porosity (decreased the cell wall thickness). As a result, the SF22 and SF15 foams reach the densification stage at a higher strain value than pure TPU foams. 

According to the summary plot in Figure \ref{img:Comp_PlotNorm}(b), SF22-20 and SF15-20 foam compressive modulus increased at a GMB volume fraction of 20\%. This is due to the stiff GMBs embedded in the cell walls increasing the stiffness of the cell walls. However, the modulus drops at 40\% volume fraction, much sooner than SF60 foams at 60\%. Due to higher particle packing density with reduced particle size, the matrix cannot transfer loads effectively between reinforcing particles at a 40\% volume percentage of smaller particles. In contrast, we can incorporate a higher volume fraction of larger GMBs in the matrix before the foam's modulus declines. When the compressive moduli are weight normalized, there is no dip for SF15-40, although the increase is not as great as in SF60-40 and SF22-40. 

% (Figure \ref{img:Comp_PlotNorm}(b))

%\begin{figure}[H]
%\centering
%\subfigure[]{
%\includegraphics[width=0.5\textwidth]{Images/Summarize.jpg}
%}
%\caption{Illustration to summarize the coupling between the print parameters and the GMB parameters.}
%\label{img:SummAll}
%\end{figure}

\begin{figure}[h!]
\centering
\subfigure[]{
\includegraphics[width=0.45\textwidth]{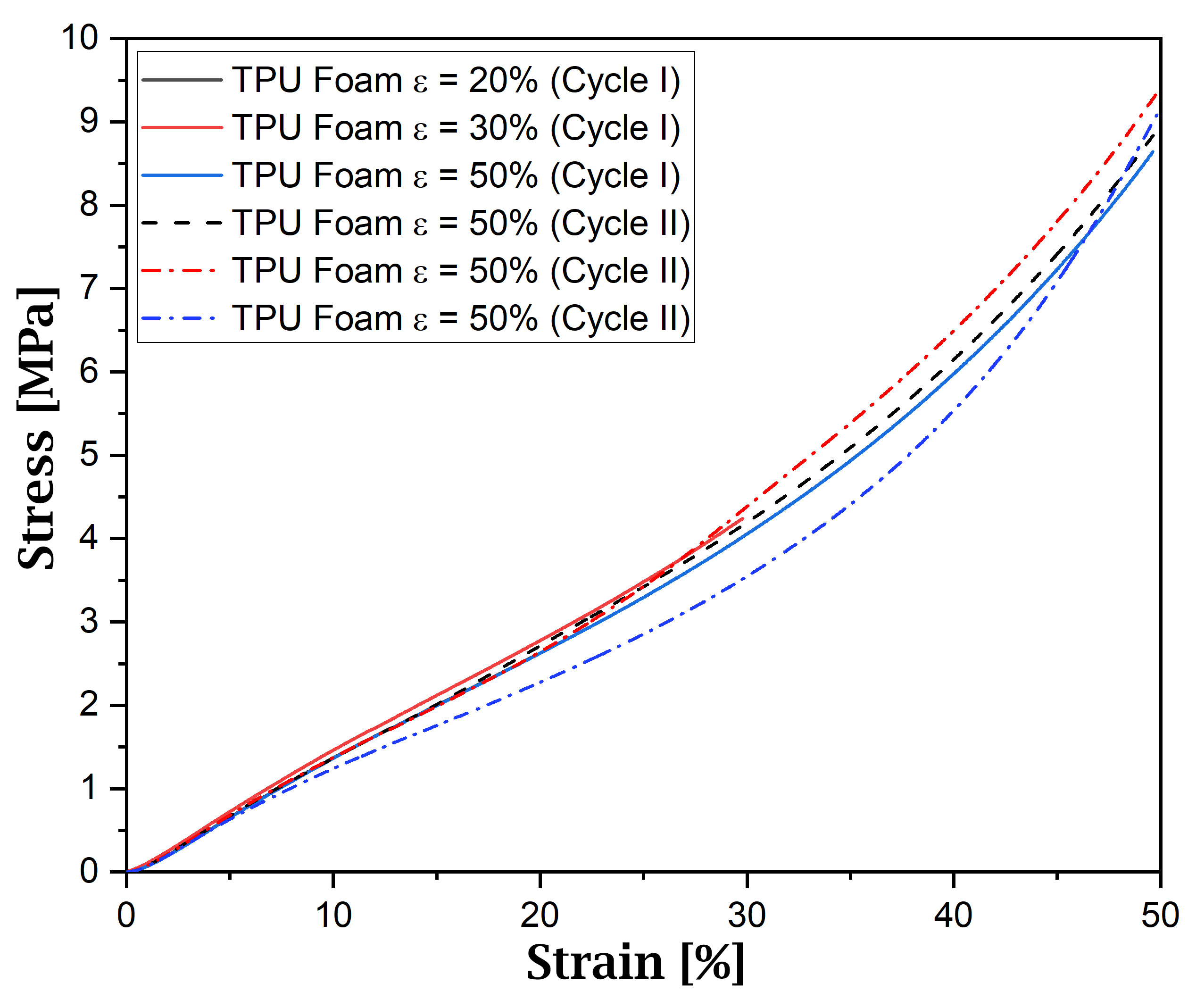}
}\hspace{3mm}
\centering
\subfigure[]{
\includegraphics[width=0.45\textwidth]{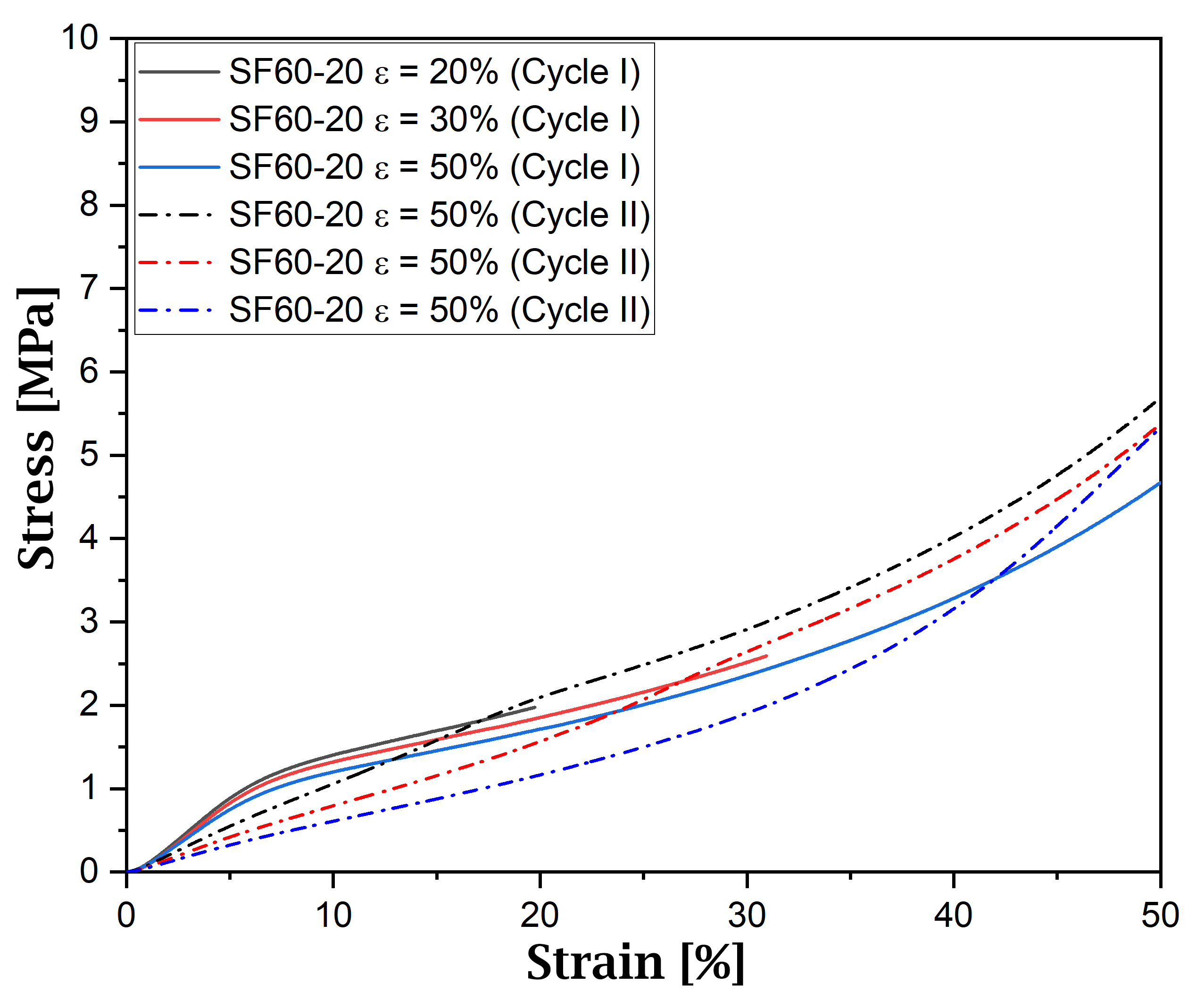}
}
\centering
\subfigure[]{
\includegraphics[width=0.45\textwidth]{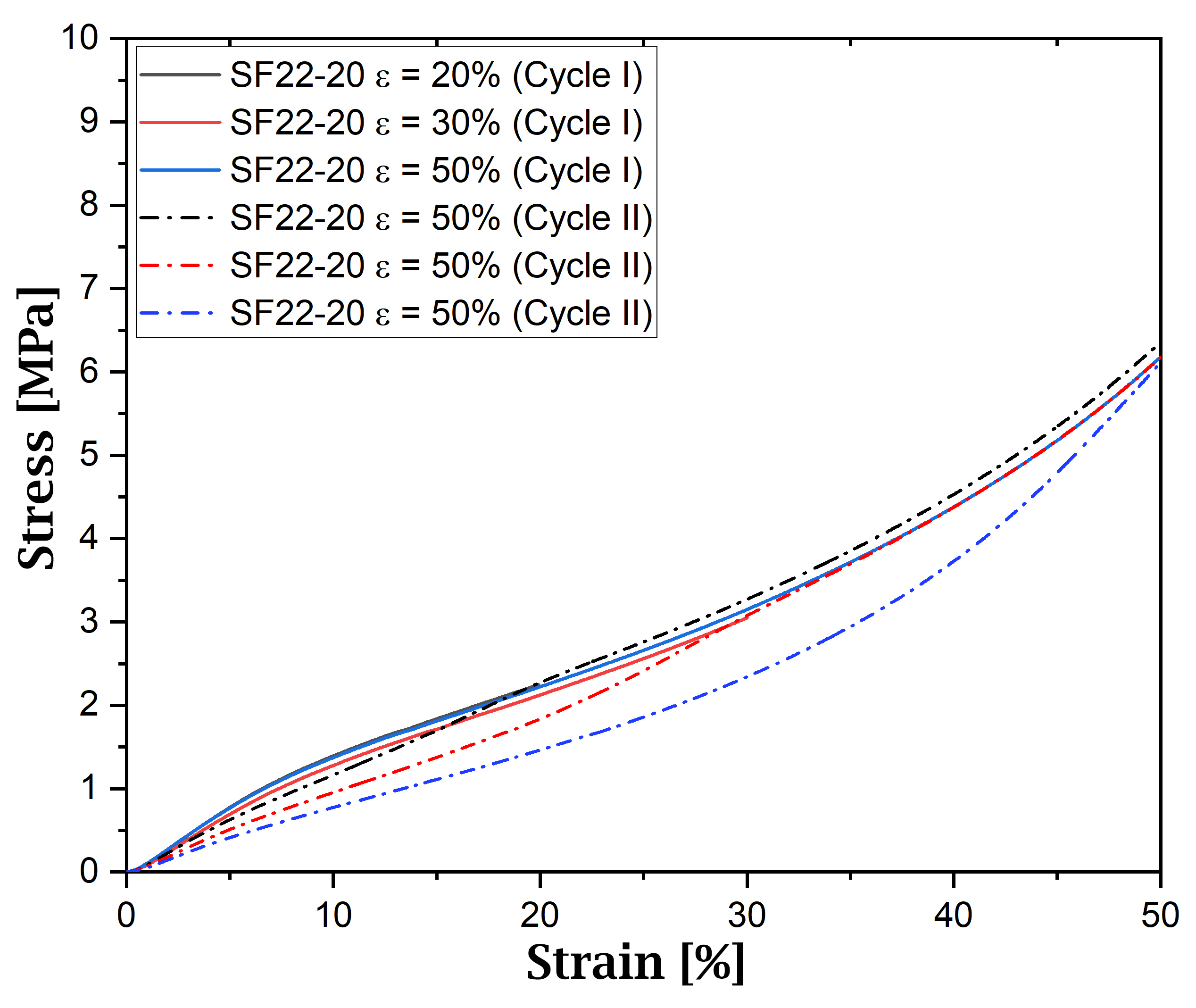}
}\hspace{3mm}
\centering
\subfigure[]{
\includegraphics[width=0.45\textwidth]{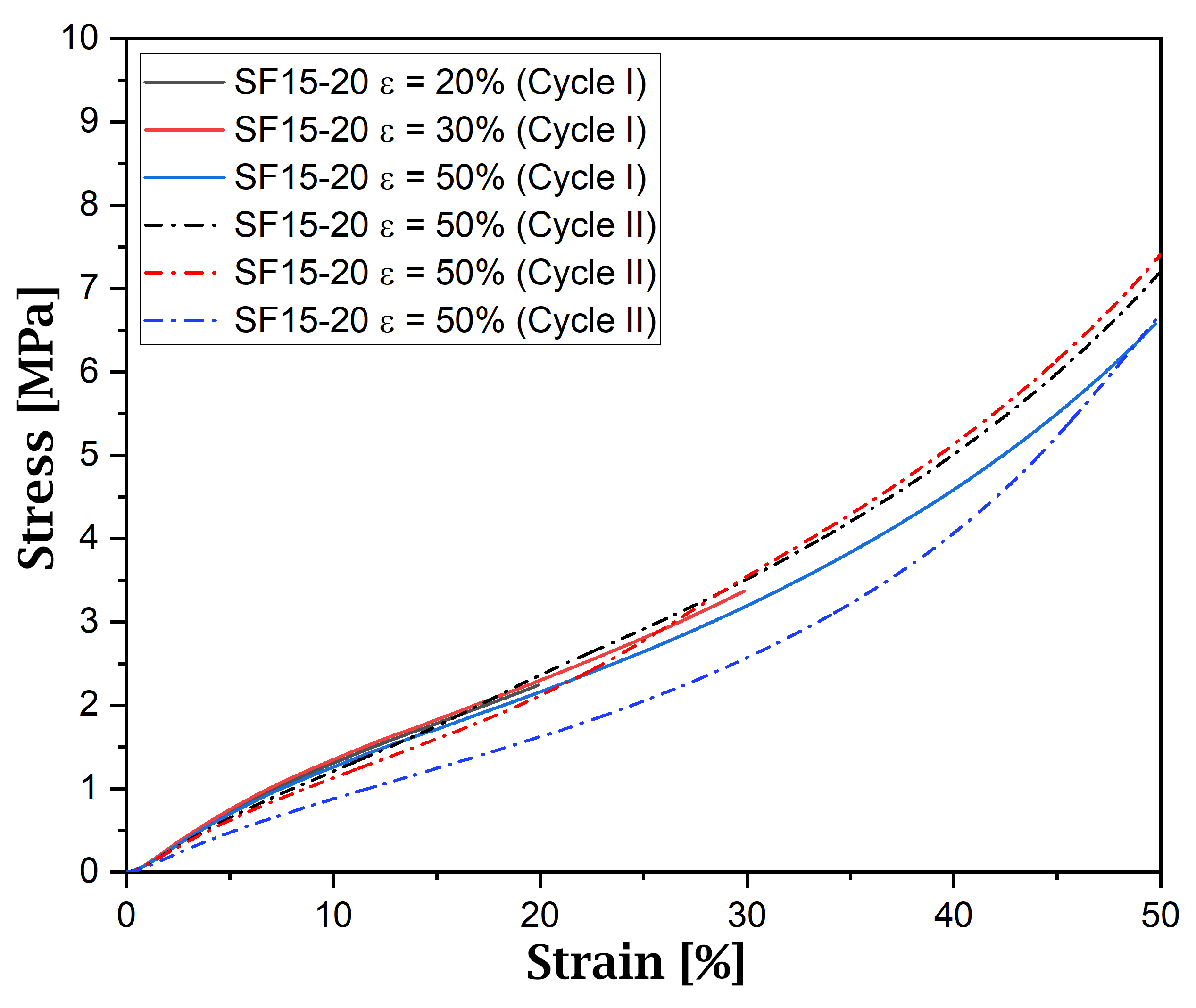}
}
\caption{Compressive stress-strain curves for (a) TPU foam, (b) SF60-20, (c) SF22-20, and (d) SF15-20 under repeated loading. Samples are compressed twice till mentioned strain values with a gap of 1 week between the loading cycles.}
\label{img:Comp_cyc}
\end{figure}

\paragraph{{\bf\em Densification Mechanics of TPU foams}}
We evaluated the residual performance of pure TPU, SF60-20, SF22-20, and SF15-20 foams under cyclic loading to evaluate the densification mechanics. In the first cycle, we subjected each set of samples to 20\%, 30\%, and 50\% strain values before allowing them to relax for one week \cite{YOUSAF2020107764}. After a one-week interval, we loaded all samples to 50\% strain values. \textbf{Figure} \ref{img:Comp_cyc} displays the representative compressive stress-strain responses of the pristine (solid lines) and compressed foams (dashed lines). Further, black, red, and blue lines represent compression to 20\%, 30\%, and 50\% strain, respectively, in the first cycle.

For pure TPU foams, both pristine and compressed foams exhibited similar compressive behavior and a slight increase in densification stress. This is because, after the first compression cycle, pure TPU foams became denser. In the second cycle after 50\% compression in the first cycle, there was a reduction in the plateau region in the TPU foam due to the excessive deformation at 50\% strain weakening the cell walls in the segregated matrix structure. The compressive response of SF60-20 showed a significant reduction in the response during the second compression cycle with maximum reduction following the 50\% compression during cycle 1. This is a result of GMBs crushing during the first compression cycle in SF60-20 foams. When the SF60-20 foams were compressed again in cycle 2, there were fewer particles to support the load and the porosity had risen due to particle crushing in the first cycle. The reduction in compressive properties of SF22-20 foams is not a result of particle compression; however, when these foams were loaded in cycle 1, the particles partially protruding from the cell walls may have been compressed against one another. During the second cycle, particles may have debonded from the matrix manifesting as a reduction in the compressive behavior. After the first cycle, the compressive properties of SF15-20 foams subjected to strain values of 20\% and 30\% in cycle 1 did not degrade. However, a substantial decrease is observed for foam that was loaded to 50\% strain in cycle 1. This is because of the reduced size of the particles in SF15-20 foams, they began to interact at higher strain values as explained in \ref{DiffGrade}, resulting in particle debonding from the TPU matrix. Therefore, we only observed a decrease in properties for SF15-20 foams when the foam was initially loaded to 50\% strain.

\begin{figure}[h!]
    \resizebox{1\textwidth}{!}{
    \begin{tabular}{|c|c|c|}
        \hline
   \textbf{SF60-20} & \textbf{SF22-20} & \textbf{SF15-20}  \\ \hline
 \includegraphics[width=0.30\textwidth]{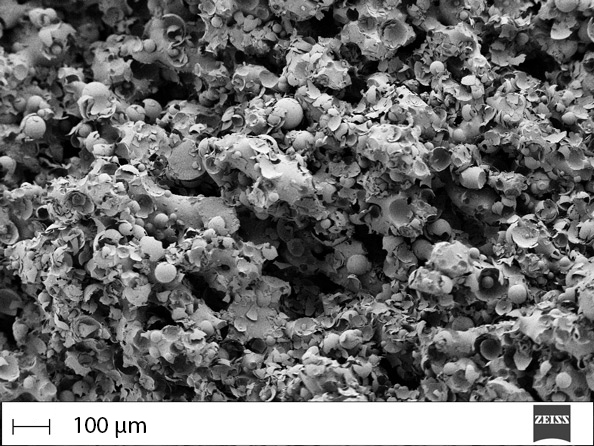} (a) &  \includegraphics[width=0.30\textwidth]{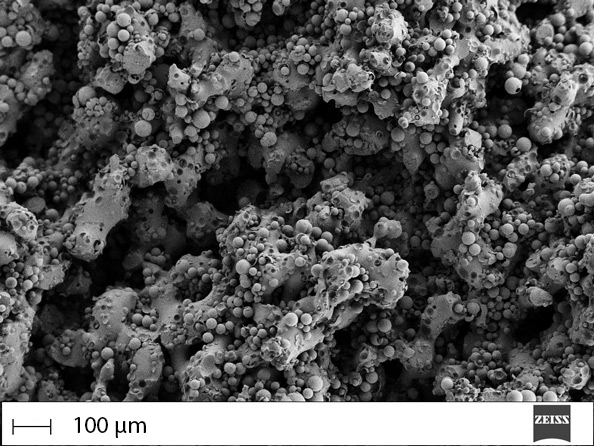} (b)
& \includegraphics[width=0.30\textwidth]{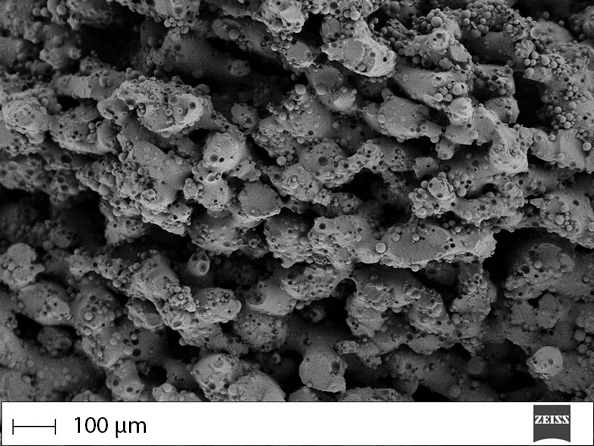} (c) \\   
        \hline
        
    \end{tabular} }
\caption{SEM images (E.H.T. = 3kV and Signal = SE2) of (a) SF60-20, (b) SF22-20, and (c) SF15-20 compressed twice to 50\% strain with a gap of 1 week between both loading cycles.}
\label{img:SEMComp_diffP}
\end{figure}

\textbf{Figure} \ref{img:SEMComp_diffP} shows the SEM morphologies of syntactic foams after 2 cycles of compression to 50\% strain as described above. We observe severe particle crushing in SF60-20 foam compared to SF22-20 and SF15-20 foams. We also observed that as the compressive strain values of cycle 1 increased, the porosity values for SF60-20 foam increased from $\sim$27\% to 43\%, due to the crushing of GMBs. However, the porosity values remained consistent for SF22-20 and SF15-20 foams, in the range of 33\% to 34\%, indicating that no particle crushing occurred during the compression cycle. The SEM morphology of twice-compressed SF22-20 foam reveals that the particles were too small to fit into the gaps, but they can still begin to interact during the initial phases of compression loading. However, due to the smaller size of GMBs in SF15-20, the particles can interact at higher strain values leading to the densification region. Although the GMBs in SF22-20 and SF15-20 foams can interact at some point during compression loading, they will not be crushed due to their exceedingly high crushing strengths.  

\begin{comment}
\begin{figure}[H]
\centering
\subfigure[]{
\includegraphics[width=0.30\textwidth]{Images/CompK20.jpg}
}
\centering
\subfigure[]{
\includegraphics[width=0.30\textwidth]{Images/CompK46.jpg}
}
\centering
\subfigure[]{
\includegraphics[width=0.30\textwidth]{Images/CompiM30k.jpg}
}
\caption{SEM images of (a) SF60-20, (b) SF22-20, and (c) SF15-20 compressed twice to 50\% strain with a gap of 1 week between both loading cycles.}
\label{img:SEMComp_diffP}
\end{figure}
\end{comment}

\begin{comment}
\begin{figure}[h!]
\centering
\subfigure[]{
\includegraphics[width=0.45\textwidth]{Images/Porosity_2.jpg}
}
\caption{Porosity values of pristine and compressed SF60-20, SF22-20, and SF15-20 foams compressed till 20\%, 30\%, and 50\% strain values}
\label{img:CompPor}
\end{figure}
\end{comment}

\paragraph{{\bf\em Process - Structure - Property Map}}

\begin{figure}[H]
    \centering
    \includegraphics[width = 0.8\textwidth]{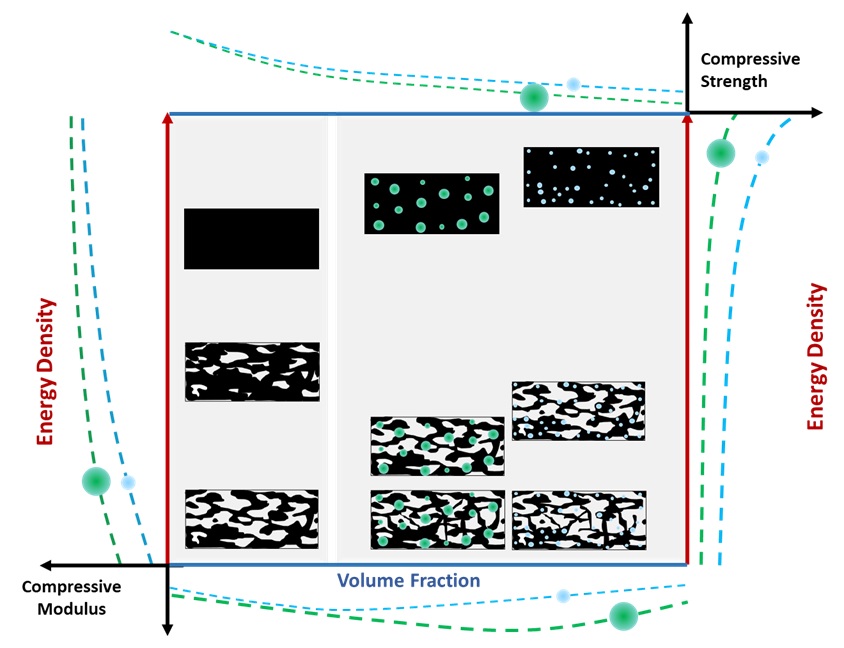}
    \caption{Proposed Structural - Process - Property map to aid with designing and manufacturing syntactic foams with segregated matrix using Selective Laser Sintering.}
    \label{img:PSPM}
\end{figure}

Based on our observations, we present a Process – Structure – Property map (\textbf{Figure} \ref{img:PSPM}) to assist with designing SLS-printed syntactic foams. This map depicts how the energy density supplied during manufacturing, GMB size and volume fraction, and associated internal micro-structure influence the foam compressive modulus and strength. We see that matrix segregation manifested during the SLS process reduces with increasing laser energy density for pure matrix. We need to supply higher energy density to syntactic foams than pure TPU to achieve a similar extent of matrix segregation. For syntactic foams, we observe that the compressive modulus of syntactic foams with segregated matrix increases with increasing volume fraction up to a certain percentage, beyond which this value drops. This optimum volume fraction reduces with reducing particle size. In addition, the compressive modulus with larger particles is higher than that with smaller particles when we have a segregated matrix. The values will converge as we approach a solid matrix with increasing energy density. On the other hand, there is a switch in compressive strength behavior. We have higher strength with smaller particles when we have a segregated matrix at lower energy densities, and the responses diverge when we approach a solid matrix.

\section{Demonstration - Compression Response of Architected Syntactic Foams}

\begin{figure}[H]
    \centering
    \includegraphics[width = 0.75\textwidth]{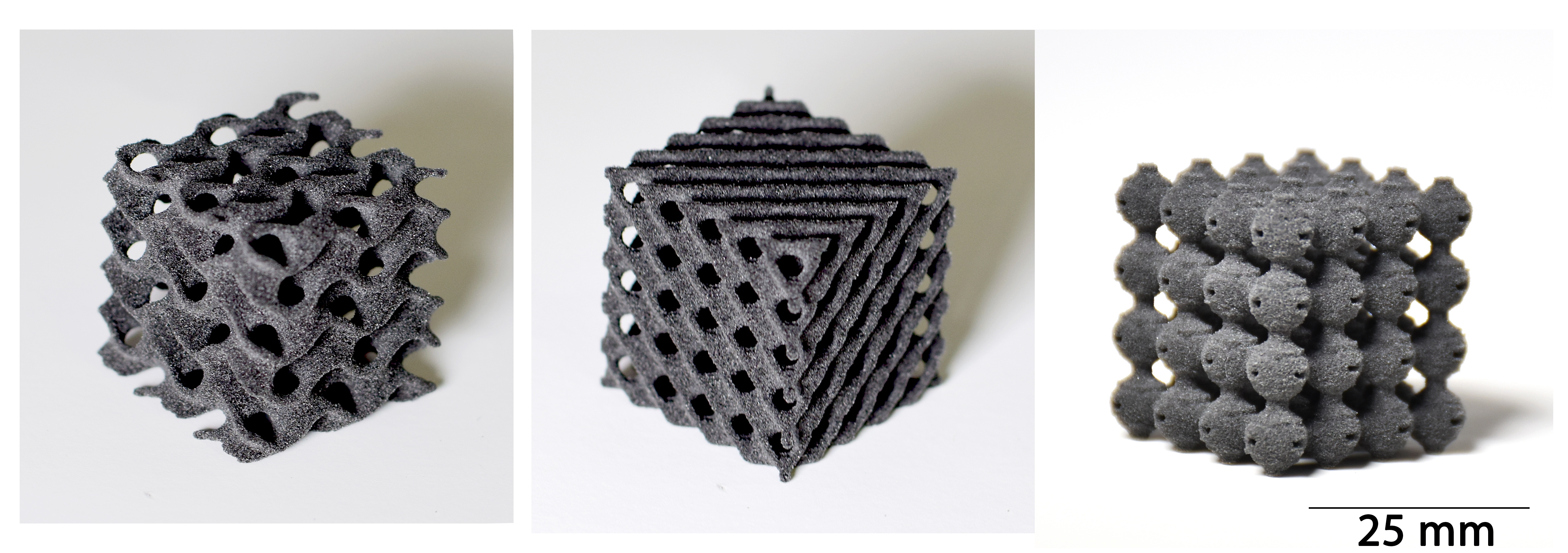}
    \caption{Printed architected SF60-40 syntactic foam samples.}
    \label{img:ArchSF}
\end{figure}

To demonstrate hierarchy at the macroscale, we manufactured architected syntactic foams with dimensions of 25mm x 25mm x 25mm. We chose three architectures, namely: i) gyroid, ii) diamond, and iii) conical as shown in \textbf{Figure} \ref{img:ArchSF}. We calculated the effective strain and effective stress by using the dimensions mentioned above. These effective properties represent the response of the overall architected structures and not of the local struts. Representative stress-strain compressive responses of all architected syntactic foams - gyroid, diamond, and conical - are summarized in \textbf{Table} \ref{tab:stiffarch} and \textbf{Table} \ref{tab:strgarch}. Observably, the gyroid and diamond architectures exhibited bending-dominated stress-strain behavior \cite{Maskery2018,Khaderi2014}, whereas the conical architecture displayed buckling-dominated (stretching-dominated) behavior \cite{Maskery2018}.

%,with two categories of architectures: (a) architectures with bending dominated response, and (b) architectures with stretching dominated response \cite{Khaderi2014,Afshar2016,Abueidda2017,Maskery2018}. 

We can observe from Table \ref{tab:stiffarch} that after incorporating GMBs into the architected designs, the stiffness increased with an increase in GMB volume fraction from 0\% to 40\% due to an expected increase in the stiffness of the struts due to the addition of GMBs. The weight normalized change increases further due to weight reduction by incorporating hollow GMBs. From Table \ref{tab:strgarch}, we see a decrease in the strength of the bending-dominated architected foams, whereas, an increase in the buckling-dominated architecture. This difference is due to the underlying mechanism of bending-dominated versus buckling-dominated compressive strengths. In bending-dominated architectures, compressive strength is dictated by the crushing of the GMBs (with low crushing strengths) due to the compression of the struts against each other. However, in buckling-dominated architectures, compressive strength is dictated by the critical buckling load of the struts which increased due to an increase in the stiffness with the addition of GMBs. \cite{ashby1983}.

\begin{table}[h]
\centering
\renewcommand{\arraystretch}{1.1}
\caption{Compressive stiffness of architected TPU and SF60-40 foams} 
\scriptsize
\resizebox{\columnwidth}{!}{%
\begin{tabular}{lccc}
\hline
{Material} & \multicolumn{3}{c}{{Stiffness (MPa)}} \\
\hline
 & \textbf{Gyroid Geometry} & \textbf{Diamond Geometry} & \textbf{Conical Geometry} \\ 
\textbf{TPU}   & 1.06 $\pm$ 0.062 & 1.46 $\pm$ 0.08  & 0.46 $\pm$ 0.012   \\ 
\textbf{SF60-40}    & 2.22 $\pm$ 0.122  & 3.03 $\pm$ 0.014  & 0.77 $\pm$ 0.083   \\ 
&  \multicolumn{2}{c}{}  &   \\ 
\% Change    & \textcolor{Green} {109.43\%} & \textcolor{Green}{107.53\%} & \textcolor{Green}{67\%}  \\ 
\% Normalized Change    & \textcolor{Green}{215.61\%} & \textcolor{Green}{193.12\%} & \textcolor{Green}{103.7\%}  \\ \hline
\end{tabular}
}
\label{tab:stiffarch}
\end{table}

\begin{table}[h]
\centering
\renewcommand{\arraystretch}{1.1}
\caption{Compressive strength of architected TPU and SF60-40 foams} 
\scriptsize
\resizebox{\columnwidth}{!}{%
\begin{tabular}{lccc}
\hline
{Material} & \multicolumn{3}{c}{{Strength (MPa)}} \\
\hline
 & \textbf{Gyroid Geometry} & \textbf{Diamond Geometry} & \textbf{Conical Geometry} \\ 
\textbf{TPU}    & 0.158 $\pm$ 0.03 & 0.23 $\pm$ 0.006  & 0.014 $\pm$ 0.0005   \\ 
\textbf{SF60-40}    & 0.11 $\pm$ 0.002 & 0.17 $\pm$ 0.005  & 0.018 $\pm$ 0.0004   \\ 
&  \multicolumn{2}{c}{}  &   \\ 
\% Change    & \textcolor{red} {-30.38\%} & \textcolor{red}{-26.09\%} & \textcolor{Green}{8.9\%}  \\ 
\% Normalized Change    & \textcolor{Green}{3.68\%} & \textcolor{Green}{6.63\%} & \textcolor{Green}{32.5\%}  \\ \hline
\end{tabular}
}
\label{tab:strgarch}
\end{table}

\section{Conclusion}\label{conc}

This paper presents a novel study on the mechanics of additively manufactured syntactic foams with segregated matrix systems. We show how additive manufacturing parameters can be coupled with GMB parameters to achieve the desired mechanical response or to tune the mechanical response of syntactic foams with the segregated matrix. To that end, this paper proposes an additive manufacturing technique for producing lightweight syntactic foams consisting of a segregated Thermoplastic Polyurethane (TPU) matrix and Glass Micro-Balloons (GMBs), which can be extended to the production of lightweight syntactic foams with intricate architectural designs. We evaluated the effect of print parameters on the mechanical response of the structures. Additionally, the effects of incorporating various grades of GMBs at various volume fractions were evaluated and discussed. The compression responses of two categories of architectures, bending-dominated and stretching-dominated, were studied for architected syntactic foams. %Finally, we revised existing theoretical micromechanical models to take the segregated matrix into account and proposed a method to link GMB and print parameters for more accurate models.

%Key conclusions from this work can be summarized as,

Our study demonstrated, for the first time, that GMB size dictates the cell wall thickness of the segregated matrix in additively manufactured polymer syntactic foams, consequently influencing porosity. Smaller GMBs in powder blends scattered more energy due to a higher particle density per area, lowering the energy absorbed by the TPU powder blend. This decreased TPU matrix cell wall thickness and increased the porosity of the foams. Therefore, blends with smaller GMBs will need more energy density (higher laser power or lower scanning speed) to compensate for a higher particle per area density than TPU/GMB blends with larger GMBs. 

Our investigation also unveiled the counter-intuitive deformation mechanics of syntactic foams with the segregated matrix associated with different GMB sizes. We revealed groundbreaking insight that different GMB parameters can be used to either enhance the stiffness or increase the densification stress. Bigger GMBs lodged inside and between the cell walls of the segregated TPU matrix created a quasi-bridge across the cell walls that increased the modulus. However, due to the lower crushing strength of larger particles, the densification stress is reduced, attributed to particle crushing. On the other hand, smaller GMBs with higher crushing strengths embedded primarily inside the cell walls manifested a matrix-dominated stress-strain response similar to pure TPU foams, with no particle crushing. These findings emphasize the importance of considering GMB sizes and grades based on the specific application to achieve the desired foam stiffness and strength. These insights also pave the way for a deeper understanding of the structural performance of syntactic foams with segregated matrix influenced by GMB parameters. 

In addition to fundamental insights from our study on syntactic foams with segregated matrix, we also demonstrated the manufacturing of stiffer and lighter foams with multi-scale architectural hierarchies. For architectures with bending-dominated deformations, we observed that GMBs can increase the compression modulus, but the compression strength is reduced. In contrast, the addition of GMBs enhanced compression modulus and strength for architectures exhibiting stretching-dominated response. 

Our findings not only demonstrate the intricate relationships between GMB and print parameters but also open up novel avenues to design and manufacture tunable syntactic foams containing segregated matrix systems.

\section*{Author Contributions}

Author H.R.T. contributed to the conceptualization of the methodology, formal analysis, investigating, and writing -- preparing the original draft. Author M.H. contributed to the manufacturing, investigating, and preparing of the draft. Authors M.T. and H.S. contributed to performing experiments and formal analysis. Author P.P. contributed to the conceptualization of the methodology, writing -- reviewing and editing, visualization, verification, supervision, project administration, and funding acquisition. 
%%%%%%%%%%%%%%%%%%%%%%%%%%%%%%%%%%%%%%%%%%
\section*{Funding}

This research was partly funded by the U.S. Department of Defense (DoD) Office of Naval Research - Young Investigator Program (ONR - YIP) Grant [N00014-19-1-2206] through the {\em{Sea-based Aviation: Structures and Materials Program}}.
%%%%%%%%%%%%%%%%%%%%%%%%%%%%%%%%%%%%%%%%%%
\section*{Acknowledgment}

This research was partially supported by the University of Wisconsin - Madison College of Engineering Shared Research Facilities and the NSF through the Materials Science Research and Engineering Center (DMR-1720415) using instrumentation provided at the UW - Madison Materials Science Center. 

%%%%%%%%%%%%%%%%%%%%%%%%%%%%%%%%%%%%%%%%%%
\section*{Data Availability}
The data that support the findings of this study are available from the corresponding author, PP, upon request.
%\section*{References}

\bibliographystyle{unsrt}
\bibliography{SLS_SF_bib}

%----------------------------------------------------------------------------------------
\clearpage
\appendix
%\section{Appendix}

\section{Selective Laser Sintering Process}\label{app:SLSprocess}

In the SLS printing process, a roller pushes a layer of powder with a specified layer height from the feed bed to the print bed. Then, the powder layer on the print bed is heated by the IR heaters to a temperature in the sintering window of the powder. Finally, a high-energy laser beam with a prescribed energy density fuses the powder to itself to form the provided 3D entity layer by layer. After the SLS printing process was complete, we carefully removed the samples and cleaned them with a sandblaster – a brush was then used to remove some of the remaining particles from the surface.
 
%\subsubsection{Laser Sintering Parameters}

The print parameters can influence the printed part's quality and mechanical response. Hence, we performed a parametric study to elucidate the impact of two key print parameters, namely laser power ratio and layer height, on the morphology and mechanical properties of printed TPU and syntactic foams. The energy supplied by the laser depends on two parameters: laser power and scanning speed. This energy from the heat supplied changes the size of the melt pool, which is directly proportional to the laser power and inversely proportional to the scanning speed. The laser energy supplied to the polymer powder is controlled by a factor called the laser power ratio in the Sinterit Lisa 3D printer \cite{Sinteritsp.zo.o.2022}. We varied the layer height and laser power ratio to determine their influence on the mechanical properties of the printed foams. First, we varied layer height as 75 $\mu$m, 125 $\mu$m, and 175 $\mu$m, for a fixed laser power ratio of 1.0. Then, the laser power ratio was chosen to be 0.75, 1.0, 1.5, and 2.0, for a fixed layer height of 125 $\mu$m. 

\begin{figure}[h!]
    \centering
    \includegraphics[width = 0.5\textwidth]{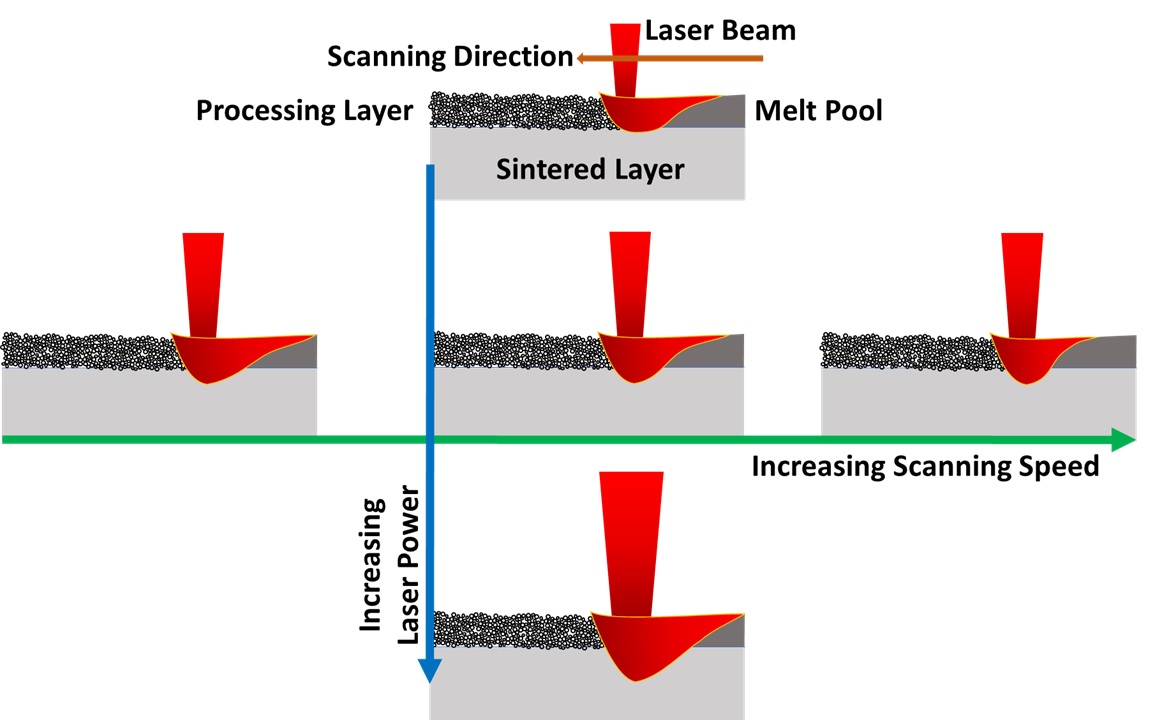}
    \caption{Illustration of the effect of Laser Power and Scanning Speed on the energy melt pool}
    \label{img:LPRIll}
\end{figure}

\section{Methods - Materials Characterization}\label{appendix_materials_charac}

\subsection{Scanning Electron Microscopy (SEM)}
Particle size distributions and microscale morphologies of TPU powder and GMBs were determined using the Zeiss Gemini 450 FESEM (3keV and SE2 signal). We spread a layer of powder on a carbon tape, observed it under the SEM, and measured the diameter using the ImageJ software \cite{Schindelin2012}. We also used SEM to understand the microstructure of the printed specimens with different parameters and to analyze the morphologies of the specimens after failure.

\subsection{Fourier Transform Infrared (FTIR) spectroscopy}
We used the attenuated total reflectance FTIR spectroscopy on the Bruker FT-IR microscope to understand the chemical compositions of the constituents to manufacture the syntactic foams. As the energy absorption ability of the powder is vital for the sintering process, we also analyzed the spectroscopic properties of the pure TPU powder and the TPU/GMB blends. Further, we used this technique to investigate the chemical changes due to the sintering process and examine the effect of print parameters on the chemical composition of the printed foams. The FTIR spectra of the syntactic foams were obtained at a resolution of 4 {$cm^{-1}$} for wavenumbers ranging from 4000 {$cm^{-1}$} to 600 {$cm^{-1}$}. 

\subsection{Differential Scanning Calorimetry (DSC)}
To evaluate the thermal properties of the TPU and determine the optimal sintering window, we used the TA Instruments QA 200 equipment. DSC allowed us to also evaluate the effect of incorporating GMBs on the thermal characteristics of the TPU/GMB blends. Approximately 8-10 mg of the sample was loaded into a Hermetic Aluminum pan, and it was rapidly heated to 225 \degree C at a rate of 20 \degree C/min to get rid of any impurities present. The sample was then cooled to -70 \degree C at a constant rate of 10 \degree C/min followed by heating to 225 \degree C at a rate of 10 \degree C/min. DSC allowed us to identify the melting temperature and the recrystallization temperature of the polymer powders, and the window between the onset of these two temperatures is the optimal sintering window. Nitrogen gas was used as a coolant and the flow rate was maintained at 50 $cm\textsuperscript{3}$/min. DSC curves were then evaluated using TA Universal Analysis Software.

\subsection{Porosity Measurements}

We used a helium porosimeter to measure the porosity values of the pure and GMB included TPU syntactic foams. The helium porosimeter consists of two cells: a chamber and a reference, with known internal volumes. We placed the foam samples inside the chamber cell for the measurements. After both cells were vacuumed until the pressure reached 0.3-0.4 psi, we loaded only the reference cell with helium gas until the pressure reached about 80-90 psi and recorded the pressure after it stabilized. Then we opened the valve connecting the chamber and reference cells to release the helium gas from the reference cell into the chamber cell. We further recorded the resulting pressure at equilibrium and used it to calculate the sample solid volume inside the chamber cell based on Boyle’s law. Solid volume can be compared with the total sample volume to obtain porosity.

For all porosity tests, we used printed foams with dimensions of 25mm x 25mm x 25mm and final print conditions. Individual porosity values of all pristine foams were obtained and compared with those of compressed foams to see the effect of compression loading. Two samples of each type of foam were chosen, and the porosity values were measured three times for each sample, with the average value chosen as the final porosity value.

\section{Results - Materials Characterization}\label{appendix_results}

\subsection{Powder size distribution} \label{PSD}

We obtained the particle size distribution with the help of ImageJ, an image analysis software \cite{Schindelin2012}. SEM images of the TPU powder and different GMBs were loaded in the ImageJ software and a measurement scale of 100 $\mu$m was used. We measured approximately 250 particles using this measurement tool. 

\begin{figure}[h!]
\centering
\subfigure[]{
\includegraphics[width=0.45\textwidth]{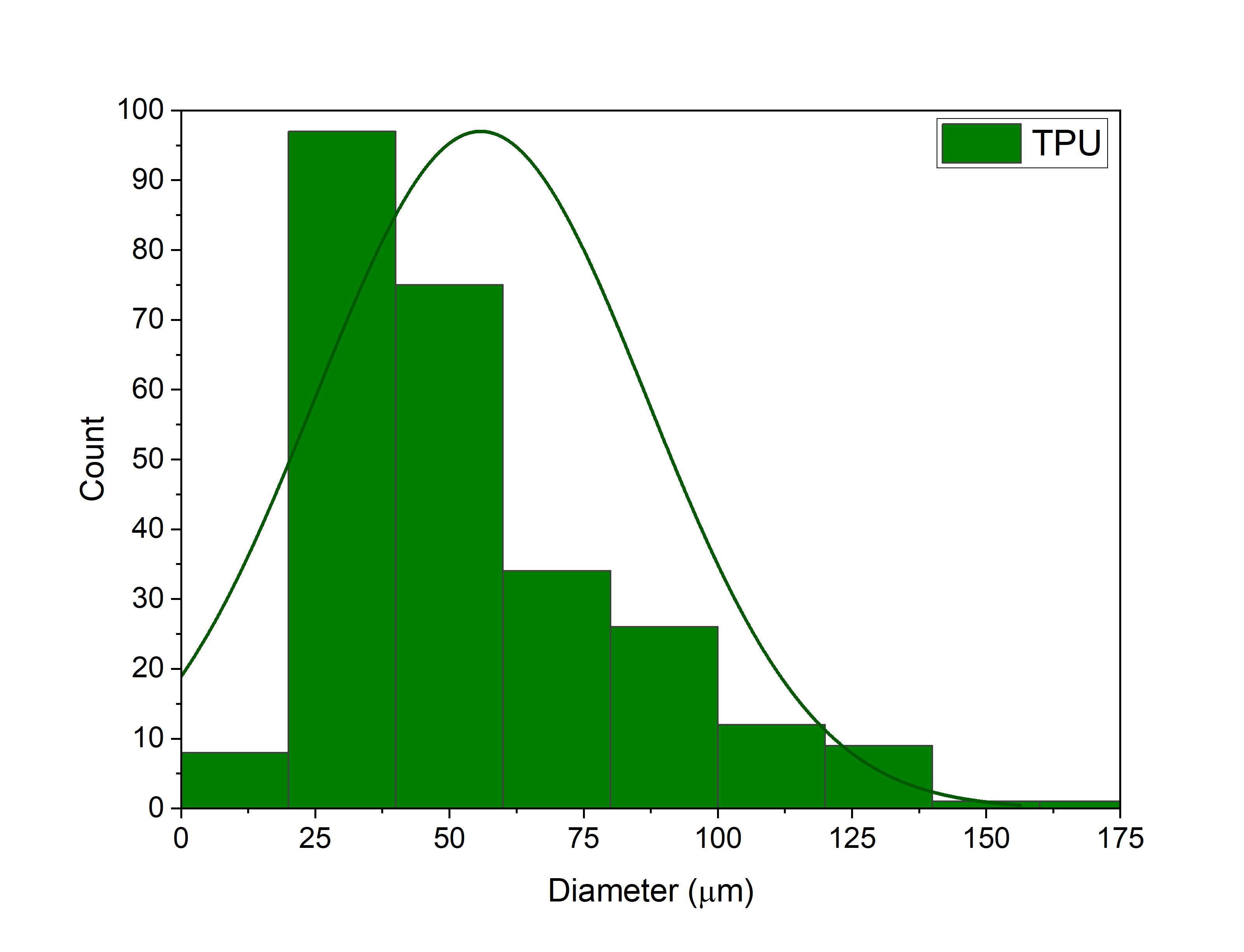}
}
\centering
\subfigure[]{
\includegraphics[width=0.45\textwidth]{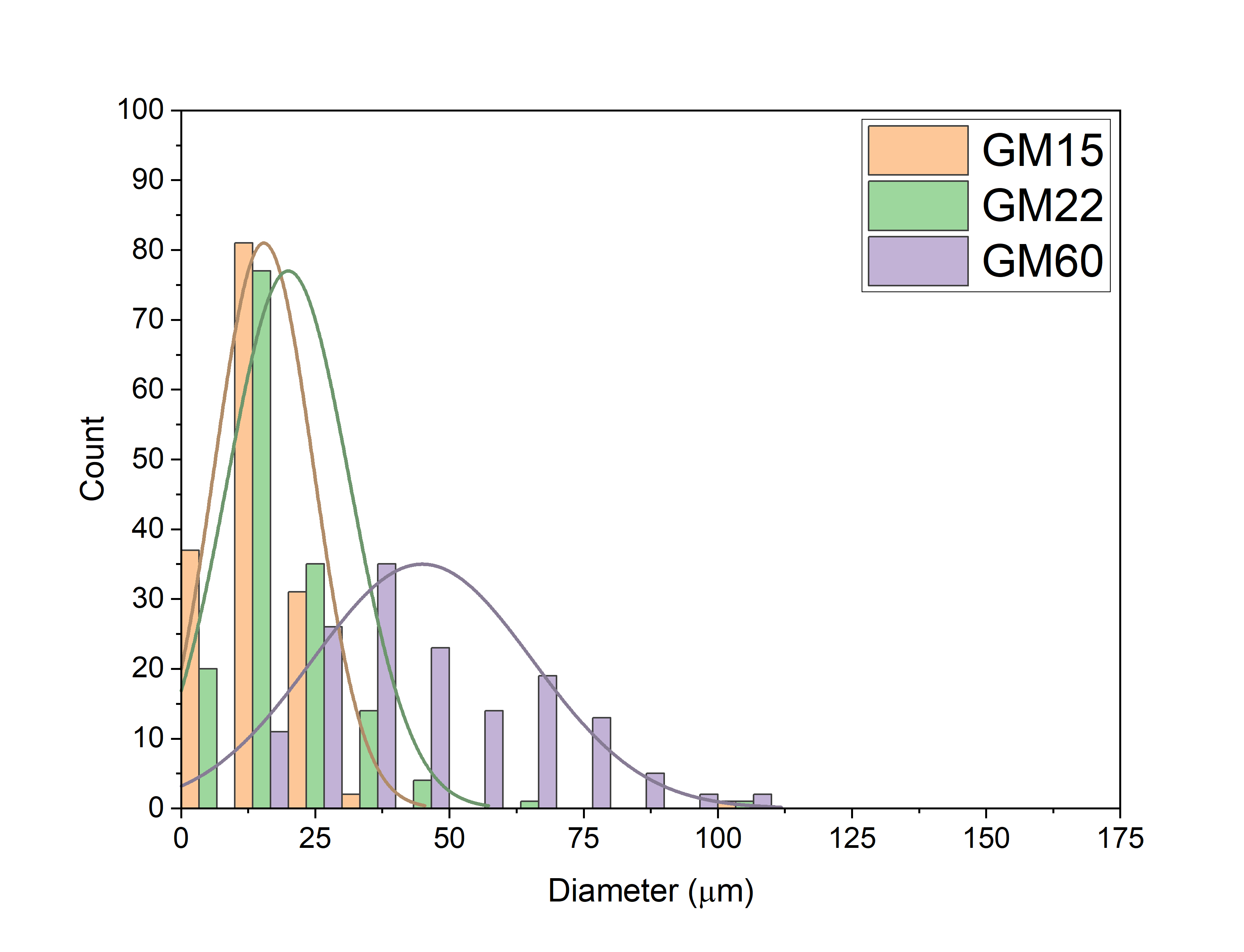}
}
\caption{Particle size distribution of (a) pure TPU powder and (b) GM60, GM22, and GM15 particles}
\label{img:partdist}
\end{figure}

Particle size distributions are shown in \textbf{Figure} \ref{img:partdist} for the TPU powder and the three grades of GMBs chosen for this study. From Figure \ref{img:partdist}(a) we can see that the diameter of the TPU powder ranged from approximately 15 $\mu$m to 140 $\mu$m. We observed from Figure \ref{img:partdist}(b), that the particle size for GM60 lay in the range of 15 $\mu$m to 120$\mu$m, which is a particle range similar to the TPU powder itself. However, for GM22 and GM15, the size distributions varied from 5 $\mu$m to 60 $\mu$m and 5 $\mu$m to 45 $\mu$m, respectively. For any volume fractions of GMBs added to the TPU powder, the effective size distribution of the mix was between the distribution of TPU and the particular grade of GMBs.  

\subsection{Thermoanalytical (DSC) Measurements}

\begin{figure}[h!]
\centering
\includegraphics[width=0.5\textwidth]{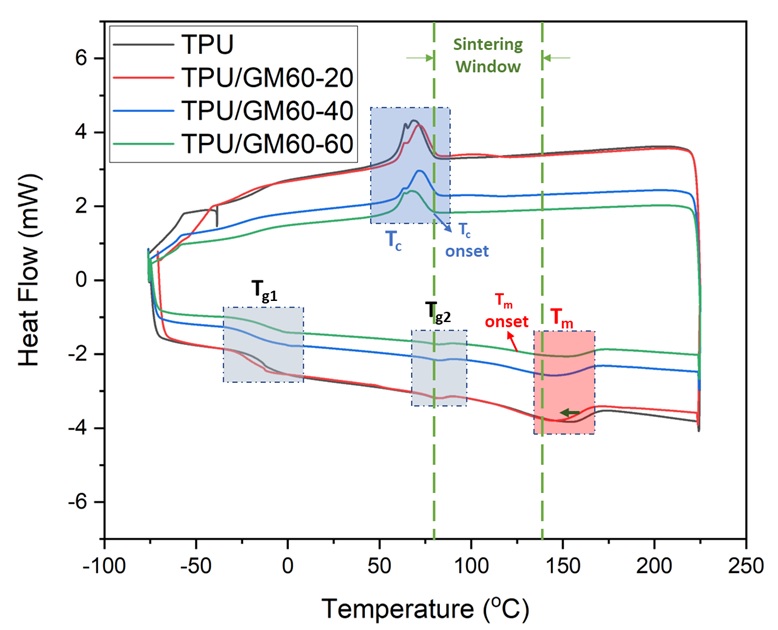}
\caption{DSC curves for the following powder blends: TPU Powder, TPU/GM60-20, TPU/GM60-40, and TPU/GM60-60; to show the effect of the addition of GMBs on the sintering window for the powder blends.}
\label{img:DSC}
\end{figure}

We used DSC measurements to determine the effect of adding different volume fractions of GMBs to TPU powder on the sintering window for the blend. To study melting, we heated pure TPU powder and GM60 mixes to 225 \degree C. For crystallization, the materials were chilled to -70 \degree C. The sintering window is the temperature difference between the melting and crystallization onset points (see \textbf{Figure} \ref{img:DSC}). The sintering window shortened from 29.88 to 26.89 \degree C as the volume percentage of GM60 GMBs grew from 0\% to 60\%. GMBs in the powder blend cannot be sintered, decreasing the sintering window. TPU has hard and soft segments, hence the DSC plot showed two glass transition temperatures and two crystallization peaks. Both glass transition temperatures, $T_{g1}$ (for soft segments) and $T_{g2}$ (for hard segments), remained within the same temperature range for pure TPU powder and the other TPU/GM60 blends. The DSC observations are summarized in \textbf{Table} \ref{tab:DSC}. 

\begin{table}[h]
\centering
\renewcommand{\arraystretch}{1.2}
\caption{Thermoanalytical measurements of powder blends with different volume fractions of GM60 in the TPU powders}
\begin{tabular}{lcccccc}
\hline
{Powder Blend}  & \textbf{$T_{g1}$ ($ ^\circ C $)}  & \textbf{$T_{g2}$ ($ ^\circ C $)} & \textbf{$T_{m}$ ($ ^\circ C $)} & \textbf{$T_{c}$ ($ ^\circ C $)} & \textbf{$\Delta$T ($T_{m}$ onset - $T_{c}$ onset) ($ ^\circ C $)} \\ \hline
TPU  & -16.96  & 62.70  & 143.21 & 68.27 & 29.88  \\ 
TPU/GM60-20 & -22.51  & 69.08  & 143.08  & 70.89 & 28.39  \\ 
TPU/GM60-40 & -19.53 & 61.82 & 142.08 & 71.26 & 27.36 \\ 
TPU/GM60-60 & -14.78 & 63.54 & 143.54 & 67.18  & 26.89  \\ \hline
\end{tabular}
\label{tab:DSC}
\end{table}

%----------------------------------------------------------------------------------------

\end{document}